\documentclass[preprint,showpacs,prd]{revtex4-1}
\usepackage{amssymb}
\usepackage{amsmath} 
\usepackage{color}
\usepackage{subeqnarray}
\usepackage{float}
\usepackage{amsfonts} 
\usepackage{caption}
\usepackage{graphicx} 
\usepackage{tikz}
\usetikzlibrary{positioning,arrows}
\usetikzlibrary{decorations.pathmorphing}
\usetikzlibrary{decorations.markings}
\usepackage{feynmp-auto} %

\begin{document}

\title{$\mathcal{O} \,(\theta)$ Feynman rules for quadrilinear gauge boson couplings in noncommutative standard model}
\author{Seyed Shams Sajadi}
\email{s.shams.sajadi@email.com}
\author{G. R. Boroun}
\email{boroun@razi.ac.ir}
\affiliation{Department of Physics, Razi University, Kermanshah, 67149, Iran}
\date{\today}

\begin{abstract}
We examine the electroweak gauge sector of noncommutative standard model and in particular, obtain the $\mathcal{O} \,(\theta)$ Feynman rules for all quadrilinear gauge boson couplings. Surprisingly, an electroweak-chromodynamics mixing appears in the gauge sector of noncommutative standard model, where photon as well as the neutral weak boson are coupled directly to three gluons. The phenomenological perspectives of the model in $W^-W^+ \rightarrow Z Z$ scattering are studied and it is shown that there is a characteristic oscillatory behaviour in azimuthal distribution of scattering cross sections which can be interpreted as a direct signal of noncommutative standard model. Assuming the integrated luminosity 100 $\mathrm{fb}^{-1}$, the number of $W^-W^+ \rightarrow Z Z$ subprocesses are estimated for some values of noncommutative scale $\Lambda_{\mathrm{NC}}$ at different center of mass energies and the results are compared with predictions of the standard model.
\end{abstract}
\maketitle


\section{Introduction}\label{intro}
Over the last 15 years, there has been an increasing interest to study noncommutative standard model as a candidate for beyond Glashaw, Weinberg and Salam model of
particle physics~\cite{sleva,ahmad,chegal,mansour,ettefagh,anaa,ohl,muller}. This is partially because of the modern foundations of string theory, where in its context it was shown that noncommutative models occur in description of low energy excitations of open strings in the presence of a constant background $B$-field~\cite{seiberg}. On the other hand, noncommutative theories are of interest on their own as a nontrivial generalization of ordinary gauge theories on a deformed background which is defined by commutation relations $[x^\mu,x^\nu] = i \theta^{\mu \nu}$, where $x^{\mu}$ denotes the spacetime 4-vector and $\theta^{\mu\nu}$ is a constant, real and antisymmetric matrix of dimensions $\mathrm{GeV}^{-2}$~\cite{snyder}. It is generally believed that signatures of noncommutative spacetime can be observed at string scale, typically on the order of Planck distance, where the quantum effects of gravitational fields become significant. Although the Planck scale ($10^{19} \, \mathrm{GeV}$) is actually far from our direct access, however, assuming the possibility of the existence of large extra dimensions and given that the onset of string effects is at TeV scale, signatures of noncommutative background are expected to be observable at a few TeV~\cite{nima1,nima2}. Today, there is a positive attitude, both experimentally and theoretically, for a new physics at TeV scale and intense experimental efforts~\cite{cms,exp1}, phenomenological studies~\cite{cakir,phe1}, as well as many independent model buildings~\cite{martin,bhowmick} are currently underway to find out signs of the new physics beyond the standard model. Noncommutative extension of the standard model appears to be a suitable candidate for the new physics and it may finally be realized by nature at TeV domain of energy. Nevertheless, the situation is uncertain and some alternative scenarios with compatible successes, such as SUSY models~\cite{susy1,susy2} and D-Branes~\cite{bran1,bran2}, also have been suggested all awaiting for experimental confirmation.

Noncommutative field theories can be constructed by the Moyal-Weyl correspondence, where the usual product of functions are promoted to an associative star product which is defined as~\cite{moyal1,moyal2}
\begin{equation}\label{eq:moyal}
f(x) \star g(x) = f(x) \exp\left(\frac{i}{2} \theta^{\mu \nu} {\frac{\overleftarrow{\partial}}{\partial x^{\mu}}}  {\frac{\overrightarrow{\partial}}{\partial y^{\nu}}}\right)g(y)|_{y=x} \,\,.
\end{equation}
There are, however, two serious problems in constructing noncommutative standard model based on this approach. The first and probably the most important difficulty in the Moyal-Weyl correspondence is the problem of charge quantization. That is, the possible charges for the matter fields are automatically restricted to the values $-1, 0, +1$. Secondly, it turns out that in the non-Abelian case, only noncommutative models with $\mathrm{U(N)}$ gauge symmetry are allowed in this approach~\cite{chaichian1,hayakawa}. An idea to resolve these problems was proposed by Chaichian et al.~\cite{chaichian2}. They built up a noncommutative $\mathrm{U(3)}\otimes \mathrm{U(2)}\otimes \mathrm{U(1)}$ gauge theory and then reduced it to noncommutative $\mathrm{SU(3)}\otimes \mathrm{SU(2)}\otimes \mathrm{U(1)}$ model by breaking the original symmetry of the theory in an appropriate manner. The model, however, introduces some extra bosons (three vector and one scalar) in comparison with the standard model. An alternative solution which cure both the problems and at the same time preserves the particle content of the standard model, is to use of Seiberg-Witten maps for noncommutative gauge field $\hat{A}_\mu$ and the corresponding gauge transformation parameter $\hat{\Lambda}$~\cite{seiberg}. Under such a construction noncommutative objects are written as an infinite series on deformation quantity $\theta^{\mu \nu}$ which then, upto an arbitrary order in $\theta^{\mu \nu}$, they can be expressed in terms of usual (commutative) fields and gauge parameters. Contrary to ordinary field theories because of the presence of $\star-$product the commutation relations of noncommutative gauge fields as well as the gauge parameters do not close to the Lie algebra of the symmetry group. This problem can be circumvented by constructing noncommutative models based on the enveloping algebra of the gauge group. This idea was proposed by Jur\v{c}o et al.~\cite{jurco1,jurco2} and used to extend the Siberg-Witten maps to non-Abelian gauge fields as well as the gauges coupled to matter fields. Along these lines, Calmet et al.~\cite{calmet} introduced the minimal noncommutative standard model and later developed it to the non-minimal extension (according to the freedom in choice of traces in the gauge sector) of the model~\cite{melic1,melic2}. The Seiberg-Witten construction by Jur\v{c}o and collaborators.~\cite{jurco1} also has found applications in relation with gravitation and topology~\cite{garcia1,aschieri}. Recently, some of geometric and topological implications of noncommutative Wilson loops have been studied in Ref.~\cite{garcia2}.

Noncommutative models have a rich phenomenological content and many interesting features. In particular, noncommutative standard model introduces new interactions which are forbidden in the standard model. Such interactions can be used to test the model through rare events (see for example~\cite{behr,buric2,xia}) and may lead to a distinct phenomenology. Another remarkable feature of the model is that, there are contributions from the Higgs part of the noncommutative action which enter directly into pure gauge sector of the theory and can affect the electroweak gauge boson interactions. The Feynman rules for Trilinear Gauge boson Couplings (TQC's) including contributions from the Higgs sector for both the minimal and non-minimal models have been already obtained~\cite{melic1}. Recently, the rules for the Higgs couplings with gauge bosons have been also completed~\cite{betabi}. Here, we are going to obtain the $\mathcal{O}\,(\theta)$ Feynman rules for Quadrilinear Gauge boson Couplings (QGC's) in both the minimal and non-minimal noncommutative standard model.

This paper is organized as follows. In Sec.~\ref{model}, we review briefly the minimum required basis of noncommutative standard model. In particular, we emphasize on the gauge and Higgs sectors of the model and identify the relevant interactions for QGC's. In Sec.~\ref{rules}, we obtain the Feynman rules for all QGC's including the anomalous couplings of photon and the weak boson $Z^0$ to three gluons. In Sec.~\ref{discuss}, we study phenomenological perspectives of the model in $W^-W^+\rightarrow ZZ$ scattering and in Sec.~\ref{conclu} summarize the paper and outline the concluding remarks. We will use a notation close to the original paper by Meli\'{c} et al.~\cite{melic1} to make the next review section short and the results readily applicable for phenomenological studies.
\section{Seiberg-Witten maps and noncommutative standard model}\label{model}
To begin, let us recall that the action of noncommutative standard model can be easily built up from the action of the standard model by replacing the normal products between fields with $\star$ ones, and the fields by their corresponding Seiberg-Witten maps. For the fermion field $\psi$ and an arbitrary gauge field $V_{\mu}$, upto the first order of deformation parameter $\theta^{\mu \nu}$ this means~\cite{jurco1,calmet}
\begin{eqnarray}\label{eq:seibergmatter}
   \psi \rightarrow \hat{\psi}[\psi,V] = \phi - \frac{1}{2} \theta^{\rho \sigma} V_{\sigma} \partial_\rho \phi + \frac{i}{8} \theta^{\rho \sigma} [V_{\rho}, V_{\sigma}],
\\
\label{eq:seiberggauge}
  V_\mu \rightarrow \hat{V_{\mu}}[V] = V_{\mu} + \frac{1}{4} \theta^{\rho \sigma} \{ \partial_\rho V_\sigma + F_{\rho \sigma} , V_\sigma \} \, .
\end{eqnarray}
A hat on letters is to indicate the noncommutative objects. The bracket $\{\,,\,\}$ denotes the anticommutator of operators and $F_{\mu \nu}$ is the usual field strength tensor. The noncommutative field tensor is defined as $\hat{F}_{\mu \nu} = \partial_\mu \hat{V}_\nu - \partial_\nu \hat{V}_\mu - ig[\hat{V}_\mu,  \hat{V}_\nu]_\star$ and $\star-$commutator means  $\hat{V}_\mu \star \hat{V}_\nu - \hat{V}_\nu \star \hat{V}_\mu$. In order to construct noncommutative standard model one can choose the gauge potential $V_{\mu} = g' \mathcal{A}_{\mu} Y + g \sum^{3}_{a=1} B^{a}_{\mu} T^{a}_{\mathrm{L}} + g_{\mathrm{s}} \sum^{8}_{b=1} G^{b}_{\mu} T^{b}_{\mathrm{S}}$, where $\mathcal{A}_{\mu}$, $B^a_{\mu}$ and $G^b_{\mu}$ represent the fields associated respectively to $\mathrm{U_{Y}(1)}$, $\mathrm{SU_{L}(2)}$ and $\mathrm{SU_{C}(3)}$ gauge groups with corresponding coupling constants $g'$, $g$, $g_{\mathrm{s}}$. Also, $Y$, $T^{a}_{\mathrm{L}}$ and $T^{b}_{\mathrm{S}}$ are generators of the relevant structure groups.

On the other hand, the noncommutative Higgs field $\hat{\Phi}$ is given by the hybrid Seiberg-Witten map as
\begin{eqnarray}\label{eq:seiberghiggs} \nonumber
 \Phi \rightarrow  \hat{\Phi}[\Phi,V,V'] & = & \Phi + \frac{1}{2} \theta^{\rho \sigma} V_{\sigma} \bigl[ \partial_\rho \Phi - \frac{i}{2} (V_{\rho} \Phi - \Phi V'_{\rho}) \bigl]
   \\
   & + &
   \frac{1}{2} \theta^{\rho \sigma} \bigl[ \partial_\rho \Phi - \frac{i}{2} (V_{\rho} \Phi - \Phi V'_{\rho}) \bigl] V'_{\sigma} \, .
\end{eqnarray}
Observe that, noncommutative Higgs field can be transformed under two different gauge groups on the left and the right corresponding respectively to gauge potentials $V_{\mu}$ and $V'_{\mu}$~\cite{melic1}. The action of noncommutative standard model can be formally written as
\begin{equation}\label{eq:NCaction}
  S_{\mathrm{NCSM}} = S_{\mathrm{ferimion}} + S_{\mathrm{gauge}} + S_{\mathrm{Higgs}} +S_{\mathrm{Yukawa}} \, ,
\end{equation}
The relevant expressions for each part of the above action have been obtained in~\cite{melic1}. For our purposes, it suffices to rewrite only the gauge and Higgs parts in detail.
\subsection{Gauge sector}
The gauge action is~\cite{melic1,melic2}
\begin{equation}\label{eq:NCgauge1}
   S_{\mathrm{g}}  =  - \frac{1}{2} \int {d^4 x  \mathrm{Tr} \frac{1}{\mathbf{G^2}} \hat{F}_{\mu \nu} \star \hat{F}^{\mu \nu} } , \frac{1}{g^2_I} =  \mathrm{Tr} \frac{1}{\mathbf{G^2}} T^a_I T^a_I ,
\end{equation}
where, $g_I$'s are usual (commutative) coupling constants $g'$, $g$, $g_{\mathrm{s}}$. Here, the trace is over all the unitary and irreducible representations of the symmetry group and $\mathbf{G}$ is an operator which commutes with generators of the gauge group and determines the coupling constants of the model. It is in general, a function of $Y$ and Casimir operators of $\mathrm{SU_{L}(2)}$ and $\mathrm{SU_{C}(3)}$. Because the noncommutative fields are valued in the enveloping algebra of the gauge group, the trace in Eq.~\eqref{eq:NCgauge1} is not unique depends strongly on the choice of a representation for gauge fields~\cite{calmet,melic1}. All the representations which appear in the standard model are important and must be considered. Using Seiberg-Witten map~\eqref{eq:seiberggauge} and the $\star-$product prescription in~\eqref{eq:moyal} upto the first order in $\theta^{\mu \nu}$ we can rewrite the gauge action as
\begin{eqnarray}\label{eq:NCgauge} \nonumber
   & S &_{\mathrm{g}}  =  - \frac{1}{2} \int {d^4 x \, \mathrm{Tr} \frac{1}{\mathbf{G^2}} F_{\mu \nu} F^{\mu \nu} }
   \\
   & + & \theta^{\rho \sigma} \int {d^4 x \, \mathrm{Tr} \frac{1}{\mathbf{G^2}} \left(\frac{1}{4} F_{\rho \sigma} F_{\mu \nu} - F_{\rho \mu} F_{\sigma \nu}\right)F^{\mu \nu} } \, .
\end{eqnarray}
\subsubsection{Minimal noncommutative standard model}
The simplest choice for representation of gauge fields is the adjoint representation. In this case the trace is taken independently over generators of the symmetry groups, i.e., respectively over $Y$, $T^{a}_{\mathrm{L}}$ and $T^{b}_{\mathrm{S}}$. In this case the resulting (gauge) action will remain as close as possible to that of the standard model. By substitution of gauge potential $V_\mu$ in~\eqref{eq:NCgauge} and rearranging the fields we get
\begin{eqnarray}\label{eq:mNCgauge} \nonumber
& S &^{\mathrm{m}}_{\mathrm{g}}  =
  - \frac{1}{2} \int d^4 x \biggl( \frac{1}{2} \mathcal{A}_{\mu \nu} \mathcal{A}^{\mu\nu} + \, B^{a}_{\mu \nu} B^{\mu \nu,a} + G^{a}_{\mu \nu} G^{\mu \nu,a} \biggl)
   \\
  & + & \frac{1}{4} g_{\mathrm{s}} d^{abc} \theta^{\rho \sigma} \int d^4 x \left( \frac{1}{4}  G^{a}_{\rho\sigma} G^{b}_{\mu\nu} - G^{a}_{\rho\mu} G^{b}_{\sigma\nu} \right)G^{\mu\nu,c} \, ,
\end{eqnarray}
where,
\begin{subequations}
\begin{eqnarray}
  \mathcal{A}_{\mu \nu} & = & \partial_{\mu}  \mathcal{A}_{\nu} - \partial_{\nu}  \mathcal{A}_{\mu} \, ,
  \\ \slabel{eq:fields1}
  B^a_{\mu \nu} & = & \partial_{\mu}  B^a_{\nu} - \partial_{\nu}  B^{a}_{\mu} + g \, \epsilon^{abc} B^{b}_{\mu} B^{c}_{\nu} \, ,
  \\ \slabel{eq:fields2}
  G^a_{\mu \nu} & = & \partial_{\mu}  G^a_{\nu} - \partial_{\nu}  G^{a}_{\mu} + g_{\mathrm{s}} f^{abc} G^{b}_{\mu} G^{c}_{\nu} \, . \slabel{eq:fields3}
\end{eqnarray}
\end{subequations}
The $\mathcal{A}_{\mu}$ and $B^a_{\mu}$ fields can be expressed in terms of physical fields as usual using
\begin{subequations}
\begin{eqnarray}
B^1_{\mu} & = & \frac{ W^+_{\mu} + W^-_{\mu}}{\sqrt{2}}\, , \,\,\,\,\,\, B^2_{\mu} =i\frac{ W^+_{\mu} - W^-_{\mu}}{\sqrt{2}} \, ,
\\ \slabel{eq:physicalfields1}
  \mathcal{A}_{\mu} & = & \cos{\theta_w} A_{\mu} - \sin{\theta_w} Z_{\mu} \, ,
  \\ \slabel{eq:physicalfields2}
  B^3_{\mu} & = & \sin{\theta_w} A_{\mu} + \cos{\theta_w} B_{\mu} \, . \slabel{eq:physicalfields3}
\end{eqnarray}
\end{subequations}
Here, $A_\mu$ is the photon field, $Z_\mu$ and $W^\pm_\mu$ are weak boson fields and $\theta_{w}$ stands for the weak mixing angle. From Eq.~\eqref{eq:mNCgauge} it follows that in the minimal noncommutative model and at leading order of $\theta^{\mu \nu}$, the electroweak part of the gauge action is the same as that of the standard model. The QCD sector, however, differs from its corresponding action in the standard model and has already been discussed in~\cite{melic2}.

By substitution of field tensors~\eqref{eq:fields1} -~\eqref{eq:fields2} in~\eqref{eq:mNCgauge} we can isolate the relevant parts of the gauge action to (electroweak) QGC's as
\begin{equation}\label{eq:SMrules}
  - \frac{1}{2} g^2 \int d^4 x \, \epsilon^{abc} \, \epsilon^{ab'c'} B^{b}_{\mu} B^{c}_{\nu} \, B^{\mu,b'} B^{\nu,c'} \, ,
\end{equation}
which then using~\eqref{eq:physicalfields1} -~\eqref{eq:physicalfields3} can be written as
\begin{eqnarray}\label{eq:SMrulesexpan}  \nonumber
   \sim \, \, \, &g^2& \int d^4 x \, \big( W^+_{\mu} W^{-\mu} W^+_{\nu} W^{-\nu} + \cdots
   \\ \nonumber
   & + & \sin^2{\theta_w} W^+_{\mu}W^{-\mu} A_{\nu} A^{\nu} + \cdots
   \\ \nonumber
   & + & \cos^2{\theta_w} W^+_{\mu}W^{-\mu} Z_{\nu} Z^{\nu} + \cdots
   \\
   & + & \sin{\theta_w} \cos{\theta_w} W^+_{\mu}W^{-\mu} A_{\nu} Z^{\nu} + \cdots
  \big).
\end{eqnarray}
where, each line represents a typical of a number of interaction terms which differ from each other in indices and upto an numerical factor. From these interactions we obtain the gauge part of Feynman rules for $W^-W^+W^-W^+$, $W^-W^+ \gamma \gamma$, $W^-W^+ ZZ$ and $W^-W^+Z\gamma$ couplings in the context of minimal model. Notice that because the minimal extension of the standard model leaves the electroweak gauge action invariant, these expressions will be the same as the rules in the standard model. The rules will be given in Sec.~\ref{rules}.
\subsubsection{Non-minimal noncommutative standard model}
In the non-minimal model, the trace in~\eqref{eq:NCgauge1} is chosen over all particles existent in the model (with different quantum numbers) which have covariant derivatives acting on them. In the standard model, there are five multiples of fermions for each generation and one Higgs multiplet (see Table I in~\cite{calmet,melic1}). The non-minimal gauge action up to the linear order in $\theta^{\mu \nu}$ will be
 \begin{eqnarray}\label{eq:nmNCgauge} \nonumber
  S^{\mathrm{nm}}_{\mathrm{g}} &=& S^{\mathrm{m}}_{\mathrm{g}} + g'^3k_{1}\theta^{\rho \sigma} \int d^4 x \left( \frac{1}{4} \mathcal{A}_{\rho \sigma} \mathcal{A}_{\mu \nu} - \mathcal{A}_{\mu \rho} \mathcal{A}_{\nu \sigma}\right) \mathcal{A}^{\mu \nu}
  \\ \nonumber
  &+& g' g^2 k_2 \theta^{\rho \sigma} \int d^4 x \biggl[\left( \frac{1}{4} \mathcal{A}_{\rho \sigma} B^a_{\mu \nu} - \mathcal{A}_{\mu \rho} B^a_{\nu \sigma} \right) B^{\mu \nu,a}
  \\ \nonumber
  & + & \mathrm{Cyclic \,\, Permutation \,\, of \,\, Fields}\biggl]
  \\ \nonumber
  &+& g' g_{\mathrm{s}}^2 k_3 \theta^{\rho \sigma} \int d^4 x \biggl[\left( \frac{1}{4} \mathcal{A}_{\rho \sigma} G^a_{\mu \nu} - \mathcal{A}_{\mu \rho} G^a_{\nu \sigma} \right) G^{\mu \nu,a}
  \\
  & + & \mathrm{Cyclic \,\, Permutation \,\, of \,\, Fields}\biggl] \,.
\end{eqnarray}
The constants $k_i$, $i=1,2,3$ are model parameters which by using a set of constraints can be determined in terms of coupling constants of the model~\cite{behr,buric2,ana2}.

The pure electroweak QGC's arise from following interactions
\begin{eqnarray}\label{eq:nmrules}  \nonumber
  g'  g^2 & k_2 &  \int d^4 x \, \epsilon^{abc}
  \theta^{\rho \sigma} \bigl[ \mathcal{A}_{\rho \sigma} \partial_{\mu} B^a_{\nu} B^{\mu,b} B^{\nu,c}
  \\ \nonumber
  & + & \mathcal{A}_{\mu \nu} \partial_{\rho} B^a_{\sigma} B^{\mu,b} B^{\nu,c}
  + \frac{1}{2} \mathcal{A}_{\mu \nu} ( B^b_{\rho} B^c_{\sigma} \partial^{\mu} B^{\nu,a}
 \\ \nonumber
  & - &  \frac{1}{2} B^b_{\rho} B^c_{\sigma} \partial^{\nu} B^{\mu,a} )
  - \mathcal{A}_{\mu \rho} (\partial_{\nu} B^a_{\sigma} B^{\mu,b} B^{\nu,c}
  \\ \nonumber
  & - &  \partial_{\sigma} B^a_{\nu} B^{\mu,b} B^{\nu,c}
  + B^b_{\nu} B^c_{\sigma} \partial^{\mu} B^{\nu,a}
  \\ \nonumber
  & - &  B^b_{\nu} B^c_{\sigma} \partial^{\nu} B^{\mu,a})
  - \mathcal{A}^{\mu \nu} (\partial_{\mu} B^a_{\rho} B^b_{\nu} B^c_{\sigma}
  \\ \nonumber
  & - & \partial_{\rho} B^a_{\mu} B^b_{\nu} B^c_{\sigma}
  + B^b_{\mu} B^c_{\rho} \partial_{\nu} B^a_{\sigma}
  - B^b_{\mu} B^c_{\rho} \partial_{\sigma} B^a_{\nu})
  \\ \nonumber
  & - &  \mathcal{A}^{\nu \sigma} (\partial_{\mu} B^a_{\rho} B^{\mu,b} B^{\nu,c}
  - \partial_{\rho} B^a_{\mu} B^{\mu,b} B^{\nu,c}
  \\
  & + & B^b_{\mu} B^c_{\rho} \partial^{\mu} B^{\nu,a}
   -   B^b_{\mu} B^c_{\rho} \partial^{\nu} B^{\mu,a})
  \bigl]
 \end{eqnarray}
By inserting~\eqref{eq:physicalfields1} -~\eqref{eq:physicalfields3} in~\eqref{eq:nmrules} we find
\begin{eqnarray}\label{eq:nmrulesphysical}  \nonumber
\sim g' g^2 & k_2 &  \int d^4 x \theta^{\rho \sigma} \bigl( \cos^2{\theta_w} \partial_{\rho} A_{\sigma} \partial_{\mu} W^-_{\nu}  W^{+\mu} Z^\nu + \cdots
  \\ \nonumber
  & + & \sin{\theta_w} \cos{\theta_w} \partial_{\rho} A_{\sigma} \partial_{\mu} W^+_{\nu}  W^{-\mu} A^\nu + \cdots
  \\
  & + & \sin^2{\theta_w} \partial_{\rho} Z_{\sigma} \partial_{\mu} W^+_{\nu}  W^{-\mu} Z^\nu + \cdots \bigl)
 \end{eqnarray}
An important point to be noted here is that the $W^-W^+W^-W^+$ coupling is not affected from interactions in the gauge sector of non-minimal noncommutative model because there is no interaction term containing four charged boson fields in~\eqref{eq:nmrulesphysical}.

On the other hand, in addition to pure elctroweak gauge couplings, because of the interactions involved in~\eqref{eq:nmNCgauge} there is also an electroweak-chromodynamics mixing in the gauge sector of non-minimal model which arises from interactions in the last two lines of~\eqref{eq:nmNCgauge}. They are
\begin{eqnarray}\label{eq:QCDrules} \nonumber
   & g'& g^3_{\mathrm{s}} f^{a b c} \int d^4 x
  \theta^{\rho \sigma} \bigl[\mathcal{A}_{\rho \sigma} \partial_{\mu} G_{\nu}^a G^{\mu,b} G^{\nu,c}
   \\ \nonumber
 & + &
 \mathcal{A}_{\mu \rho} ( \partial_{\nu} G_{\sigma}^a G^{\mu,b} G^{\nu,c}
 - \partial_{\sigma} G_{\nu}^a G^{\mu,b} G^{\nu,c}
  \\ \nonumber
 & + &
 \partial^{\mu} G^{\nu,a} G_{\nu}^{b} G_{\sigma}^{c}
 - \partial^{\nu} G^{\mu,a} G_{\nu}^{b} G_{\sigma}^{c} )
  \\ \nonumber
 & + &
 \mathcal{A}_{\mu \nu} ( \partial^{\mu} G_{\rho}^a G^{\nu,b} G_{\sigma}^{c}
 - \partial_{\rho} G^{\mu,a} G^{\nu,b} G_{\sigma}^{c}
  \\ \nonumber
 & + &
  \partial^{\nu} G_{\sigma}^a G^{\mu,b} G_{\rho}^{c}
- \partial_{\nu} G^{\sigma,a} G^{\mu,b} G_{\rho}^{c} )
 \\ \nonumber
 & + &
 \mathcal{A}_{\nu \sigma} ( \partial_{\mu} G_{\rho}^a G^{\mu,b} G^{\nu,c}
 - \partial_{\rho} G_{\mu}^a G^{\mu,b} G^{\nu,c}
 \\
 & + &
 \partial^{\mu} G^{\nu,a} G_{\mu}^b G_{\rho}^c
- \partial^{\nu} G^{\mu,a} G_{\mu}^b G_{\rho}^c ) \bigl] \, .
\end{eqnarray}
Here $a,b,c$ run from 1 to 8 for gluon fields $G_\mu$. Proceeding as before, the relevant interactions for the coupling of photon to gluons will be
\begin{eqnarray}\label{eq:nmrulesqcd} \nonumber
\sim & g'& g^3_{\mathrm{s}} f^{a b c}  \int d^4 x \theta^{\rho \sigma} \bigl[ \cos{\theta_w} \partial_{\rho} A_{\sigma} \partial_{\mu} G_{\nu}^a G^{\mu,b} G^{\nu,c}
\\
& + & \cos{\theta_w} (\partial_{\mu} A_{\rho} \partial_{\mu} G_{\sigma}^a G^{\mu,b} G^{\nu,c} + \cdots )
 ]
 \end{eqnarray}
Also, for $Z^0$ coupling to gluons we obtain
\begin{eqnarray}\label{eq:nmrulesqcd2} \nonumber
\sim & g'& g^3_{\mathrm{s}} f^{a b c}  \int d^4 x \theta^{\rho \sigma} \bigl[ \cos{\theta_w} \partial_{\rho} A_{\sigma} \partial_{\mu} G_{\nu}^a G^{\mu,b} G^{\nu,c}
\\
& + & \cos{\theta_w} (\partial_{\mu} A_{\rho} \partial_{\mu} G_{\sigma}^a G^{\mu,b} G^{\nu,c} + \cdots )
 ]
 \end{eqnarray}
\subsection{Higgs sector}
The Higgs part of noncommutative action is
\begin{eqnarray}\label{eq:higgsaction} \nonumber
  S_{ \, \mathrm{Higgs}} &=& \int d^4 x \bigl[(\hat{D}_{\mu}\hat{\Phi})^{\dagger} \star (\hat{D}^{\mu} \hat{\Phi}) - \, \mu^2 \hat{\Phi}^{\dagger} \star \hat{\Phi}
  \\
  & - & \lambda \, \hat{\Phi}^{\dagger} \star \hat{\Phi} \star \hat{\Phi}^{\dagger} \star \hat{\Phi}\bigl].
\end{eqnarray}
Here, $\mu$ and $\lambda$ are respectively the mass parameter and coupling constant. Also, $\hat{D}_{\mu}$ denotes the covariant derivative which is defined for noncommutative Higgs field as $\hat{D}_{\mu} = \partial_{\mu} \hat{\Phi}- i \hat{V}_{\mu} \star \hat{\Phi} + i \hat{\Phi} \star \hat{V'}_{\mu}$.
The expansion of~\eqref{eq:higgsaction} using~\eqref{eq:moyal},~\eqref{eq:seiberghiggs} yields
\begin{eqnarray}\label{eq:higgsaction2} \nonumber
  S_{ \, \mathrm{Higgs}} &=& \int d^4 x \bigl[(\mathbf{D}_{\mu}\Phi)^{\dagger} (\mathbf{D}^{\mu} \Phi) - \, \mu^2 \Phi^{\dagger}  \Phi
  - \lambda (\Phi^{\dagger} \Phi)^2\bigl]
  \\ \nonumber
  &+& \frac{1}{2} \theta^{\mu \nu} \int d^4 x \Phi^{\dagger} \bigl[ \mathbf{U}_{\mu \nu} + \mathbf{U}^{\dagger}_{\mu \nu} +  \frac{1}{2} \mu^2 F_{\mu \nu}
   \\
   & - & 2i \lambda \Phi(\mathbf{D}_{\mu} \Phi)^{\dagger} \mathbf{D}_{\nu} \Phi \bigl] \Phi \, .
\end{eqnarray}
where, $\mathbf{D}_{\mu} = \partial_{\mu} - i \, \mathbf{V}_{\mu}$ and the operator $\mathbf{U}_{\mu \nu}$ is
\begin{eqnarray}\label{eq:uoperator} \nonumber
   \mathbf{U}_{\mu \nu} &=& \bigl[ \overleftarrow{\partial}^\varrho + i \mathbf{V}^\varrho \bigl]
   \bigl[ -\partial_\varrho \mathbf{V}_\mu \partial_\nu - \mathbf{V}_\mu \partial_\varrho \partial_\nu + \partial_\mu \mathbf{V}_\varrho \partial_\nu
   \\ \nonumber
   & + & i \, \mathbf{V}_\varrho \mathbf{V}_\mu \partial_\nu
    + \frac{i}{2} \mathbf{V}_\mu \mathbf{V}_\nu \partial_\varrho
    + \frac{i}{2} \partial_\varrho(\mathbf{V}_\mu \mathbf{V}_\nu)
     \\
     & + & \frac{1}{2} \mathbf{V}_\varrho \mathbf{V}_\mu \mathbf{V}_\nu + \frac{i}{2} \{\mathbf{V}_\mu , \partial_\nu \mathbf{V}_\varrho +
    \mathbf{F}_{\nu \varrho}\}
   \bigl] \, .
\end{eqnarray}
Here, $\mathbf{V}_{\mu}$ is a $2\times2$ matrix which is defined as $\mathbf{V}_{\mu} = g' \mathcal{A}_{\mu} Y_{\mathrm{\Phi}} \mathbf{1} + g \, B^{a}_{\mu} T^{a}_{\mathrm{L}}$. The explicit form of $\mathbf{V}_{\mu}$ is~\cite{melic1}
\begin{equation}\label{eq:matrixv}
 \mathbf{V}_{\mu} =
\begin{pmatrix}
e A_{\mu} + g\frac{(1 - 2 \sin^2{\theta_w})}{2\cos{\theta_w}} Z_{\mu}  & \frac{g}{\sqrt{2}} W^{+}_{\mu}& \\
\frac{g}{\sqrt{2}} W^{-}_{\mu} & -\frac{g}{2\cos{\theta_w}} Z_{\mu}  &
\end{pmatrix}
\end{equation}
Analysis of Eq.~\eqref{eq:higgsaction2} reveals that the Higgs sector induce contributions into pure gauge sector of the noncommutative standard model. Proceeding similarly as in~\cite{melic1} the interactions yielding to QGC's are those terms in ~\eqref{eq:higgsaction2} which contain multiplication of four $\mathbf{V}_{\mu}$ matrices, i.e.,
\begin{eqnarray}\label{eq:higgsrules} \nonumber
  - \frac{1}{2} & \theta^{\mu \nu} & \int d^4 x  \, \Phi^{\dag}  \bigl( i \mathbf{V}_{\varrho} \mathbf{V}^{\varrho} \mathbf{V}_{\mu} \mathbf{V}_{\nu}
+ i \mathbf{V}_{\varrho} \mathbf{V}^{\mu} \mathbf{V}_{\nu} \mathbf{V}^{\varrho}
 \\ \nonumber
& - &
 i \mathbf{V}_{\varrho} \mathbf{V}^{\mu} \mathbf{V}^{\varrho} \mathbf{V}_{\nu}
+
 i \mathbf{V}_{\mu} \mathbf{V}_{\nu} \mathbf{V}_{\varrho} \mathbf{V}^{\varrho}
-
 i \mathbf{V}_{\mu} \mathbf{V}_{\varrho} \mathbf{V}_{\nu} \mathbf{V}^{\varrho}
 \\
 & + & \mathrm{Hermitian \, Conjugate}
  \bigl) \Phi \, ,
\end{eqnarray}
Using the explicit form of $\mathbf{V}_{\mu}$ and choosing the Higgs field to be in unitary gauge
\begin{equation}\label{eq:higgsvac}
 \Phi(x) = \frac{1}{\sqrt{2}}
\begin{pmatrix}
0 & \\
h(x) + \upsilon &
\end{pmatrix} \, , \quad\quad \upsilon = \sqrt{-\mu^2/\lambda}
\end{equation}
after symmetry breaking, the Higgs induced interactions into pure gauge sector are found to be
\begin{eqnarray}\label{eq:higgs22} \nonumber
  - i\frac{2M^2_W}{g} & \theta^{\mu \nu} & \int  d^4 x  \bigl( Z_{\rho} W^{-\rho} W^+_{\mu} Z_{\nu}  + \cdots
   \\ \nonumber
  & + &
  W^-_{\rho}  A^{\rho}  A_{\mu}  W^+_{\nu}  + \cdots
  \\ \nonumber
  & + &
  W^-_{\rho}  A^{\rho}  Z_{\mu}  W^+_{\nu}  + \cdots
  \\
  & + &
  W^-_{\rho}  W^{+\rho}  W^{-\mu}  W^{+\nu} + \cdots \bigl) \, ,
\end{eqnarray}
where, we have used $\upsilon = \frac{2M_W}{g}$ for later convenience. These interactions are of dimension-4 and hence momentum independent. Notice that the electroweak gauge action of minimal noncommutative model is exactlly the same as that of the standard model. In this case, only the Higgs induced interactions can contribute to $\mathcal{O}(\theta)$ Feynman rules for electroweak QGC's.
\section{Feynman rules for QGC's}\label{rules}
The Feynman rule associated to a coupling diagram can be obtained straightforwardly by variation of the corresponding interactions. We used Eq.~\eqref{eq:SMrulesexpan} to obtain the standard rules for $W^-W^+W^-W^+$, $W^-W^+ \gamma \gamma$, $W^-W^+ZZ$ and $W^-W^+ Z \gamma$. In the minimal model the effective interactions which contribute to $\mathcal{O}(\theta)$ Feynman rules arise from the Higgs induced interactions, i.e.,~\eqref{eq:higgs22}. In the non-minimal model, contrary to the minimal case, the effective interactions arise from the $\mathcal{O}(\theta)$ extension of the pure gauge action. We used Eq.~\eqref{eq:nmrulesphysical} to get the rules of non-minimal model for pure electroweak QGC's and Eqs.~\eqref{eq:nmrulesqcd},~\eqref{eq:nmrulesqcd2} to derive the associated vertex functions for $\gamma ggg$ and $Z ggg$ couplings. All momenta are assumed to be incoming into vertices. Here are the $\theta-$expanded Feynman rules for all QGC's in electroweak gauge sector of noncommutative standard model.
\begin{enumerate}
%
\item $\,\,$ $W^{-} W^{+} W^{-} W^{+}$ coupling,
  \\
  \\
   \tikzset{
particle/.style={decorate, draw=black, decoration={snake=coil}},
}
\begin{tikzpicture}
[node distance=1.0cm and 1.5cm]
\coordinate[label=left:${W^{+}_\nu,p^+}$] (left up);
\coordinate[below right=of left up] (mid);
\coordinate[above right=of mid,label=right:${W^{-}_\kappa,{p'}^-}$] (right up);
\coordinate[below left=of mid,label=left:${W^{-}_\mu,{p}^{-}}$] (left down);
\coordinate[below right=of mid,label=right:${W^{+}_\lambda,{p'}^{+}}$] (right down);
\coordinate[above=0.8cm of mid] (tem1);
\coordinate[left=0.8cm of tem1] (mom1i);
\coordinate[right=0.8cm of tem1] (mom2i);
\coordinate[above=0.35cm of mid] (tem2);
\coordinate[left=0.2cm of tem2] (mom1f);
\coordinate[right=0.2cm of tem2] (mom2f);
\coordinate[below=0.8cm of mid] (tem11);
\coordinate[left=0.8cm of tem11] (mom1ii);
\coordinate[right=0.8cm of tem11] (mom2ii);
\coordinate[below=0.35cm of mid] (tem22);
\coordinate[left=0.2cm of tem22] (mom1f1);
\coordinate[right=0.2cm of tem22] (mom2f2);
\draw[>=latex,->] (mom1ii) -- (mom1f1);
\draw[>=latex,->] (mom2ii) -- (mom2f2);
\draw[>=latex,->] (mom1i) -- (mom1f);
\draw[>=latex,->] (mom2i) -- (mom2f);
\draw[particle] (left up) -- (mid);
\draw[particle] (mid) -- (right up);
\draw[particle] (left down) -- (mid);
\draw[particle] (mid) -- (right down);
\end{tikzpicture}
\begin{enumerate}
\item Standard model
\begin{eqnarray} \label{eq:SMWWWW}
 i {g}^{2} \left( 2\, g_{\nu \lambda} g_{\mu \kappa} - g_{\nu \kappa} g_{\lambda \mu}
- g_{\mu \nu} g_{\kappa \lambda} \right) \, ,
\end{eqnarray}
\item Minimal/Non-minimal model
\begin{eqnarray} \label{eq:mWWWW}
\mathrm{Eq}.~\eqref{eq:SMWWWW}
 -  \frac{3}{4} M^2_W {g}^{2} \left( \theta_{\kappa\lambda} g_{\mu\nu} + \theta_{\kappa\nu} g_{\mu\lambda} + \theta_{\mu\lambda} g_{\nu\kappa} + \theta_{\mu\nu} g_{\kappa\lambda}
 \right).
\end{eqnarray}
\end{enumerate}
%
%
Equation~\eqref{eq:SMWWWW} is the standard Feynman rule for $W^-W^+W^-W^+$ coupling. It is derived from the gauge action of the standard model and is symmetric under substitutions $\mu \rightleftarrows \kappa$ and independently under $\nu \rightleftarrows \lambda$. It is also symmetric under simultaneous substitutions of $\mu \rightleftarrows \kappa$ and $\nu \rightleftarrows \lambda$. These symmetries can be inferred from the above coupling diagram because exchanging two $W^-$ bosons with each other does not change the physical situation. A similar argument can be made also for exchange of $W^+$ bosons. The $\theta$-dependent part of the rule~\eqref{eq:mWWWW} results from the Higgs induced interactions~\eqref{eq:higgs22} in the minimal case. It satisfies explicitly the symmetries of Eq.~\eqref{eq:SMWWWW}. As we mentioned earlier, the gauge sector of non-minimal model does not contribute to $W^-W^+W^-W^+$ coupling and expression~\eqref{eq:mWWWW} is the $\theta$-expanded Feynman rule for both the minimal and non-minimal models.
%
%
\item $\,\,$ $W^{-} W^{+} \gamma \gamma$ coupling,
  \\
  \\
 \tikzset{
particle/.style={decorate, draw=black, decoration={snake=coil}},
}
\begin{tikzpicture}
[node distance=1.0cm and 1.5cm]
\coordinate[label=left:${A_\nu,q}$] (left up);
\coordinate[below right=of left up] (mid);
\coordinate[above right=of mid,label=right:${W^{+}_\kappa,{p}^+}$] (right up);
\coordinate[below left=of mid,label=left:${W^{-}_\mu,{p}^-}$] (left down);
\coordinate[below right=of mid,label=right:${A_\lambda,{q'}}$] (right down);
\coordinate[above=0.8cm of mid] (tem1);
\coordinate[left=0.8cm of tem1] (mom1i);
\coordinate[right=0.8cm of tem1] (mom2i);
\coordinate[above=0.35cm of mid] (tem2);
\coordinate[left=0.2cm of tem2] (mom1f);
\coordinate[right=0.2cm of tem2] (mom2f);
\coordinate[below=0.8cm of mid] (tem11);
\coordinate[left=0.8cm of tem11] (mom1ii);
\coordinate[right=0.8cm of tem11] (mom2ii);
\coordinate[below=0.35cm of mid] (tem22);
\coordinate[left=0.2cm of tem22] (mom1f1);
\coordinate[right=0.2cm of tem22] (mom2f2);
\draw[>=latex,->] (mom1ii) -- (mom1f1);
\draw[>=latex,->] (mom2ii) -- (mom2f2);
\draw[>=latex,->] (mom1i) -- (mom1f);
\draw[>=latex,->] (mom2i) -- (mom2f);
\draw[particle] (left up) -- (mid);
\draw[particle] (mid) -- (right up);
\draw[particle] (left down) -- (mid);
\draw[particle] (mid) -- (right down);
\end{tikzpicture}
\begin{enumerate}
\item Standard model
\begin{equation}\label{eq:SMWWAA}
-i e^2 \left( 2\, g_{\nu \lambda} g_{\mu \kappa} - g_{\nu \kappa} g_{\lambda \mu}
- g_{\mu \nu} g_{\kappa \lambda} \right) \, ,
\end{equation}
\item Minimal model
\begin{equation}\label{eq:mWWAA}
 \mathrm{Eq}.\eqref{eq:SMWWAA} - 2 M^2_W e^{2} \left( \theta_{\kappa\lambda}g_{
\mu\nu}-\theta_{\nu\kappa}g_{\mu\lambda} \right) \, ,
\end{equation}
\item Non-minimal model
\begin{eqnarray}\label{eq:nmWWAA} \nonumber
& \mathrm{Eq}&.~\eqref{eq:mWWAA}
+ g' g^2 k_2 \, \sin{\theta_w} \cos{\theta_w} \bigl( \{ ( -2 \theta_{{\alpha\rho}}g_{{\mu\kappa}}{g}_{{\nu\lambda}}+2 \theta_{{\alpha\rho}}g_{{\kappa\lambda}}g_{{\mu\nu}} ) q'_{{\rho}}
\\ \nonumber
& + &
 [( - \theta_{{\sigma\alpha}}p^{+}_{{\sigma}}+ \theta_{{
\sigma\alpha}}p^{-}_{{\sigma}} ) g_{{\mu\lambda}}
 +  2 \theta_{{\mu\alpha}}q'_{{\lambda}} +  2 \theta_{{\alpha\lambda
}}q'_{\mu}- \theta_{{\mu\alpha}}p^{-}_{{\lambda}}
- \theta_{{\alpha\lambda}}p^{+}_{{\mu}}
\\ \nonumber
& + &
 ( - p^{+}_{{
\alpha}}-2 q'_{{\alpha}}- p^{-}_{{\alpha}} ) \theta_
{{\mu\lambda}} ] g_{{\nu\kappa}}
 + [  \theta_{{\sigma\alpha}}p^{+}_{{\sigma}}g_{{\mu\nu}}-2 \theta_{{
\alpha\nu}}q'_{{\mu}}-2 \theta_{{\mu\alpha}}q'_{{\nu}}+ \theta_{{\alpha\nu}}p^{+}_{{\mu}}
+ \theta_{{\mu\alpha}}{p^{+}}_{{\nu}}
\\ \nonumber
& + &
 ( 2 q'_{\alpha}- p^{+}_{{\alpha}} ) \theta_{{\mu\nu}} ] g_{{\kappa\lambda}} + [ - \theta_{{\sigma\alpha}}p^{-}_{{\sigma}}g_{{\nu
\lambda}}+2 \theta_{{\alpha\nu}}q'_{{\lambda}}-2 \theta_{{
\alpha\lambda}}q'_{{\nu}}+ \theta_{{\alpha\lambda}}p^{-}_
{{\nu}}- \theta_{{\alpha\nu}}p^{-}_{{\lambda}}
\\ \nonumber
& + & ( - {
p^{-}}_{{\alpha}}+2 q'_{\alpha} ) \theta_{{\nu\lambda}}
 ] g_{{\mu\kappa}} + [ - \theta_{{\alpha\kappa}}
p^{-}_{{\nu}}+ \theta_{{\alpha\nu}}p^{-}_{{\kappa}} +  \theta
_{{\alpha\kappa}}p^{+}_{{\nu}}- \theta_{{\alpha\nu}}p^{+}_{{\kappa}}
\\ \nonumber
& + &
 (  p^{-}_{{\alpha}}- p^{+}_{{\alpha}}) \theta_{{\nu\kappa}} ] g_{{\mu\lambda}} + [ 2 \theta_{{\alpha\kappa}}q'_{{\mu}}+ \theta_{{\mu
\alpha}}p^{-}_{{\kappa}}- \theta_{{\alpha\kappa}}p^{+}_{{\mu}}- \theta_{{\mu\alpha}}p^{+}_{{\kappa}}
\\ \nonumber
& + &
 (  p^{+}_{{
\alpha}}+ p^{-}_{{\alpha}}+2 q'_{{\alpha}} ) \theta_
{{\mu\kappa}} ] g_{{\nu\lambda}} +  [ -2 \theta_{{
\alpha\kappa}}q'_{{\lambda}}- \theta_{{\alpha\lambda}}{
p^{-}}_{{\kappa}}+ \theta_{{\alpha\kappa}}p^{-}_{{\lambda}}+
\theta_{{\alpha\lambda}}p^{+}_{{\kappa}}
\\ \nonumber
& + &
 (  p^{+}_{{
\alpha}}+ p^{-}_{{\alpha}}+2 q'_{{\alpha}} ) \theta_
{{\kappa\lambda}} ] g_{{\mu\nu}} \} q_{{\alpha}}
 + \{  [ ( - \theta_{{\sigma\rho}}p^{+}_{{\sigma}}
+  \theta_{{\sigma\rho}}p^{-}_{{\sigma}} ) { g}_{{\mu\lambda}}
 - \theta_{{\mu\rho}}p^{-}_{{\lambda}}- \theta_{{\mu\lambda}}p^{-}_{{\rho}}
\\ \nonumber
& + &
2 \theta_{{\mu\rho}}q_{{\lambda}} - \theta_{{\rho\lambda}}p^{+}_{{\mu}}- \theta_{{\mu\lambda}}p^{+}_{{\rho}}+2 \theta_{{\rho\lambda}}q_{{\mu}} ] g_{{\nu\kappa}} +  (  \theta_{{\nu\rho}}{
p^{+}}_{{\mu}}+ \theta_{{\mu\rho}}p^{-}_{{\nu}}+ \theta_{{\mu\nu}}p^{+}_{{\rho}}- \theta_{{\mu\rho}}p^{+}_{{\nu}}
\\ \nonumber
& - & 2 \theta_{
{\nu\rho}}q_{{\mu}}+ \theta_{{\mu\nu}}p^{-}_{{\rho}}-
\theta_{{\sigma\rho}}p^{-}_{{\sigma}}g_{{\mu\nu}}) g_{{\kappa\lambda}}
+  [  \theta_{{\sigma\rho}}p^{+}_{{\sigma}}g_{{\nu\lambda}}+ \theta_{{\nu\lambda}}p^{+}_{{\rho}}+ \theta_{{\nu\lambda}}p^{-}_\rho+2
\theta_{{\nu\rho}}q_{{\lambda}}
\\ \nonumber
& - & \theta_{{\nu\rho}} p^{-}_\lambda
 +  ( - p^{-}_\nu + p^{+}_\nu )\theta_{{\rho\lambda}} ] g_{{\mu\kappa}} + [\theta_{{\nu\kappa}}p^{+}_{{\rho}}- \theta_{{\nu\kappa}}{ p^{-}}_{{\rho}}+ (  p^{-}_{{\kappa}}- p^{+}_{{\kappa}}) \theta_{{\nu\rho}}
 \\ \nonumber
 & + &
  ( - p^{+}_{{\nu}}+ p^{-}_{{\nu}} ) \theta_{{\rho\kappa}}]g_{{\mu\lambda}} +  (  \theta_{{\rho\kappa}}p^{+}_{{\mu}}- \theta_{{\mu
\kappa}}p^{+}_{{\rho}}+ \theta_{{\mu\rho}}p^{+}_{{\kappa}}-2 \theta_{{\mu\rho}}q_{{\kappa}} -  2 \theta_{{\rho\kappa}}{q}_{{\mu}} )g_{{\nu\lambda}}
 \\ \nonumber
 & + &
  ( - \theta_{{\rho\kappa}}p^{-}_{{\lambda}}+ \theta_{{\rho\lambda}}p^{-}_
{{\kappa}}+2 \theta_{{\rho\kappa}}q_{{\lambda}}-2 \theta_{{\rho\lambda}}q_{{\kappa}}- \theta_{{\kappa\lambda}}{ p^{-}}_{{\rho}} ) g_{{\mu\nu}} \} q'_{{\rho}}
+[  ( \theta_{{\mu\rho}}q'_{{\lambda}}+\theta_{{\mu\rho}}q_{{\lambda}}
\\ \nonumber
& + &
 \theta_{{\rho\lambda}}q_{{\mu}}+\theta_{{\rho\lambda}}q'_{{\mu}} ) p^{+}_{{\rho}} + ( - \theta_{{\sigma\mu}}p^{-}_{{\sigma}}-\theta_{{\sigma
\mu}}p^{+}_{{\sigma}} ) q'_{{\lambda}} +  (\theta_{{\rho\lambda}}q_{{\mu}}+ \theta_{{\rho\lambda}}{q'}_{{\mu}} ) p^{-}_{{\rho}} + ( - \theta_{{\sigma
\mu}}p^{-}_{{\sigma}}
\\ \nonumber
& - &
\theta_{{\sigma\mu}}p^{+}_{{\sigma}}) q_{{\lambda}} ] g_{{\nu\kappa}}
+ [  ( \theta_{{\nu\rho}}q'_{{\mu}}-\theta_{{\mu\rho}
}q'_{{\nu}}+\theta_{{\nu\rho}}q_{{\mu}} ) { p^{+}}_{{\rho}}
 + \theta_{{\nu\rho}}p^{-}_{{\rho}}q'_{{\mu}} + (-\theta_{{\sigma\nu}}p^{-}_{{\sigma}}
 \\ \nonumber
 & - &
 \theta_{{\sigma\nu}}p^{+}_{{\sigma}} ) q'_{{\mu}}-\theta_{{\sigma\nu}}{
 p^{+}}_{{\sigma}}q_{{\mu}} +  ( \theta_{{\sigma\mu}}{
p^{+}}_{{\sigma}}+ \theta_{{\sigma\mu}}p^{-}_{{\sigma}} ) {q'}_{{\nu}} ] g_{{\kappa\lambda}} + [  ( - \theta_{{\rho\lambda}}q'_{{\nu}}-\theta_{{\nu\rho}}{ q'}_{{\lambda}} )p^{+}_{{\rho}}
\\ \nonumber
& + &
 ( -\theta_{{\nu\rho}}{ p^{-}}_{{\rho}} + \theta_{{\sigma\nu}}p^{-}_{{\sigma}}+\theta_{{\sigma\nu}}p^{+}_{{\sigma}} ) q'_{{\lambda}} +  ( -\theta_{{\nu\rho}}q_{{\lambda}}- \theta_{{\rho\lambda}}{
 q'}_{{\nu}} ) p^{-}_{{\rho}}+\theta_{{\sigma\nu}}{ p^{-}}_{{\sigma}}q_{{\lambda}} ] g_{{\mu\kappa}}
\\ \nonumber
 & + &
[  ( -\theta_{{\nu\rho}}q_{{\kappa}}+ \theta_{{
\rho\kappa}}q'_{{\nu}} ) p^{+}_{{\rho}}+ ( \theta_{{\sigma\nu}}p^{+}_{{\sigma}}+\theta_{{\nu\rho}}p^{-}_{{\rho}}-\theta_{{\sigma\nu}}p^{-}_{{\sigma}}
) q_{{\kappa}} - \theta_{{\rho\kappa}}p^{-}_{{\rho}}q'_{{\nu}} ]g_{{\mu\lambda}}
 \\ \nonumber
 & + &
 [  ( -\theta_{{\mu\rho}}q_{{\kappa}}- \theta_{{\rho\kappa}}q'_{{\mu}}- \theta_{{\rho\kappa}}q_{{\mu}} ) p^{+}_{{\rho}}+ ( \theta_{{
\sigma\mu}}p^{+}_{{\sigma}}+ \theta_{{\sigma\mu}}p^{-}_{{
\sigma}} ) q_{{\kappa}}- \theta_{{\rho\kappa}}{ p^{-}}
_{{\rho}}q_{{\mu}} ] g_{{\nu\lambda}}
\\ \nonumber
& + &
 ( \theta_{{\rho\kappa}}q_{{\lambda}} -
 \theta_{{\rho\lambda}}q_{{\kappa}} ) g_{{\mu\nu}}p^{+}_{{\rho}}+
 (  \theta_{{\rho\kappa}}p^{-}_{{\rho}}q'_{{\lambda}}
- \theta_{{\rho\lambda}}p^{-}_{{\rho}}q_{{\kappa}}+
\theta_{{\rho\kappa}}p^{-}_{{\rho}}q_{{\lambda}} ) {
 g\_}_{{\mu\nu}} +  (  \theta_{{\mu\lambda}}p^{+}_{{\nu}}
  \\ \nonumber
 & + &
 \theta_{{\mu\nu}}p^{-}_{{\lambda}}+ \theta_{{\nu\lambda}}{
 p^{+}}_{{\mu}}+ \theta_{{\mu\lambda}}p^{-}_{{\nu}} - 2 \theta_{{\mu\nu}}q'_{{\lambda}} - 2 \theta_{{\nu\lambda}}q'_{{\mu}}+2 \theta_{{\mu\lambda}}q'_{{\nu}} ) q_{{\kappa}}
 +  ( - \theta_{{\mu\nu}}p^{-}_{{\kappa}}
 \\ \nonumber
 & - &
  \theta_{{\mu\nu}}p^{+}_{{\kappa}}- \theta_{{\nu\kappa}}p^{+}_{{\mu}}+2
\theta_{{\nu\kappa}}q_{{\mu}}- \theta_{{\mu\kappa}}{ p^{-}}_{{\nu}}+ \theta_{{\mu\kappa}}p^{+}_{{\nu}} ) q'_{{\lambda}}
 +  ( - \theta_{{\mu\nu}}p^{-}_{{\kappa}} -  2 \theta_
{{\mu\kappa}}q'_{{\nu}} +   \theta_{{\mu\nu}}p^{+}_{{\kappa}}
\\ \nonumber
& - &
 \theta_{{\mu\kappa}}p^{-}_{{\nu}}- \theta_{{\mu\kappa}}{
 p^{+}}_{{\nu}}+ \theta_{{\nu\kappa}}p^{+}_{{\mu}}-2 \theta_{{
\nu\kappa}}q'_{{\mu}} ) q_{{\lambda}} +  ( -
\theta_{{\kappa\lambda}}p^{+}_{{\nu}}- \theta_{{\nu\lambda}}{
 p^{-}}_{{\kappa}}+ \theta_{{\nu\kappa}}p^{-}_{{\lambda}}+
\theta_{{\kappa\lambda}}p^{-}_{{\nu}}
\\ \nonumber
& - &
 \theta_{{\nu\lambda}}{
 p^{+}}_{{\kappa}} ) q'_{{\mu}} +  ( - \theta_{{\nu
\kappa}}p^{-}_{{\lambda}}+ \theta_{{\nu\lambda}}p^{-}_{{
\kappa}}- \theta_{{\kappa\lambda}}p^{-}_{{\nu}}-2 \theta_{{
\kappa\lambda}}q'{{\nu}}- \theta_{{\kappa\lambda}}p^{+}_
{{\nu}}- \theta_{{\nu\lambda}}p^{+}_{{\kappa}} ) q_{{\mu}}
\\
& + &
 ( \theta_{{\mu\lambda}}p^{+}_{{\kappa}}+ \theta_{{\kappa\lambda}}p^{+}_{{\mu}}+ \theta_{{\mu\kappa}}p^{-}_{{\lambda}}+ \theta_{{\mu\lambda}}p^{-}_{{\kappa}} ) {q'}_{{\nu}} \bigl) \, .
\end{eqnarray}
\end{enumerate}
%
%
Equation~\eqref{eq:SMWWAA} for $W^-W^+ \gamma \gamma$  coupling is familiar from the standard model. The exchange of two photons would lead to a topologically equivalent diagram. The associated rules are therefore required to be symmetric under substitutions $\nu \rightleftarrows \lambda$. Equations~\eqref{eq:SMWWAA} and~\eqref{eq:mWWAA} obviously satisfy this requirement. The $\mathcal{O}(\theta)$ contribution of the rule~\eqref{eq:mWWAA} is derived from Higgs sector induced interactions. Equation~\eqref{eq:nmWWAA} represents the Feynman rule for the non-minimal extended model and contains a lengthy momentum dependent part. These terms are derived from~\eqref{eq:nmrulesphysical}. In this case, momenta must be simultaneously r with each other as $\nu, \lambda$ indices are substituted.
%
%
\item $\,\,$ $W^{-} W^{+} ZZ$ coupling,
   \\
   \\
   \tikzset{
particle/.style={decorate, draw=black, decoration={snake=coil}},
}
\begin{tikzpicture}
[node distance=1.0cm and 1.5cm]
\coordinate[label=left:${Z_\nu,k}$] (left up);
\coordinate[below right=of left up] (mid);
\coordinate[above right=of mid,label=right:${W^{+}_\kappa,{p}^+}$] (right up);
\coordinate[below left=of mid,label=left:${W^{-}_\mu,{p}^{-}}$] (left down);
\coordinate[below right=of mid,label=right:${Z_\lambda,{k'}}$] (right down);
\coordinate[above=0.8cm of mid] (tem1);
\coordinate[left=0.8cm of tem1] (mom1i);
\coordinate[right=0.8cm of tem1] (mom2i);
\coordinate[above=0.35cm of mid] (tem2);
\coordinate[left=0.2cm of tem2] (mom1f);
\coordinate[right=0.2cm of tem2] (mom2f);
\coordinate[below=0.8cm of mid] (tem11);
\coordinate[left=0.8cm of tem11] (mom1ii);
\coordinate[right=0.8cm of tem11] (mom2ii);
\coordinate[below=0.35cm of mid] (tem22);
\coordinate[left=0.2cm of tem22] (mom1f1);
\coordinate[right=0.2cm of tem22] (mom2f2);
\draw[>=latex,->] (mom1ii) -- (mom1f1);
\draw[>=latex,->] (mom2ii) -- (mom2f2);
\draw[>=latex,->] (mom1i) -- (mom1f);
\draw[>=latex,->] (mom2i) -- (mom2f);
\draw[particle] (left up) -- (mid);
\draw[particle] (mid) -- (right up);
\draw[particle] (left down) -- (mid);
\draw[particle] (mid) -- (right down);
\end{tikzpicture}
\begin{enumerate}
\item Standard model
\begin{equation} \label{eq:SMWWZZ}
-i g^2 \cos^2 {\theta_{w}} \left( 2\, g_{\nu \lambda} g_{\mu \kappa} - g_{\nu \kappa} g_{\lambda \mu}
- g_{\mu \nu} g_{\kappa \lambda} \right) \, ,
\end{equation}
\item Minimal model
\begin{eqnarray} \label{eq:mWWZZ} \nonumber
& \mathrm{Eq}&.\eqref{eq:SMWWZZ} \,\, - \, \frac{M^2_W}{\cos^2{\theta_{w}}} g^2
 \bigl[
-7 \theta_{\mu \kappa} g_{\nu \lambda} -2 \theta_{\kappa \lambda} g_{\mu \nu}
+ 2 \theta_{\nu \kappa} g_{\mu \lambda} + 2 \theta_{\mu \lambda} g_{\nu \kappa}
+ 2 \theta_{\mu \nu} g_{\kappa \lambda}
\\
& + &  \sin^2{\theta_{w}} (-3 \theta_{\kappa \lambda} g_{\mu \nu} + 3 \theta_{\nu \kappa} g_{\mu \lambda} - \theta_{\mu \lambda} g_{\nu \kappa} - \theta_{\mu \nu} g_{\kappa \lambda} + 4 \theta_{\mu \kappa} g_{\nu \lambda})
 +  \sin^4{\theta_{w}} (4 \theta_{\kappa \lambda} g_{\mu \nu} - 4 \theta_{\nu \kappa} g_{\mu \lambda})
 \bigl] \, ,
\end{eqnarray}
\item Non-minimal model
\begin{eqnarray} \label{eq:nmWWZZ} \nonumber
& \mathrm{Eq}&.\eqref{eq:mWWZZ} \, \,  + g' g^2 k_2 \sin{\theta_{w}} \cos{\theta_{w}} \bigl( \{  ( \, - \, 2 \theta_{{\alpha\rho}}g_{{\mu\kappa}}{g}_{{\nu\lambda}}+2\theta_{{\alpha\rho}}g_{{\kappa
\lambda}}g_{{\mu\nu}} ) k'_{{\rho}}
\\ \nonumber
& + &
 [ ( - \theta_{{\sigma\alpha}}p^{+}_{{\sigma}} + \theta_{{
\sigma\alpha}}p^{-}_{{\sigma}} ) g_{{\mu\lambda}}
+ 2 \theta_{{\mu\alpha}}k'_{{\lambda}} +  2 \theta_{{\alpha\lambda}}k'_{{\mu}}
 -  \theta_{{\mu\alpha}}p^{-}_{{\lambda}} - \theta_{{\alpha\lambda}}p^{+}_{{\mu}}
\\ \nonumber
& + &
 ( - p^{+}_{{\alpha}}-2 k'_{{\alpha}}- p^{-}_{{\alpha}} ) \theta_
{{\mu\lambda}} ] g_{{\nu\kappa}} + [  \theta_{{
\sigma\alpha}}p^{+}_{{\sigma}}g_{{\mu\nu}}-2 \theta_{{\alpha\nu}}k'_{{\mu}}-2 \theta_{{\mu\alpha}}k'_{{\nu}}+\theta_{{\alpha\nu}}p^{+}_{{\mu}}+ \theta_{{\mu\alpha}}{p^{+}}_{{\nu}}
\\ \nonumber
& + &
 ( 2 k'_{{\alpha}}- p^{+}_{{\alpha}}) \theta_{{\mu\nu}} ] g_{{\kappa\lambda}} + [ - \theta_{{\sigma\alpha}}p^{-}_{{\sigma}}g_{{\nu
\lambda}}+2 \theta_{{\alpha\nu}}k'_{{\lambda}}-2 \theta_{{\alpha\lambda}}k'_{{\nu}}+ \theta_{{\alpha\lambda}}p^{-}_{{\nu}}- \theta_{{\alpha\nu}}p^{-}_{{\lambda}}
\\ \nonumber
& + &
 ( - {p^{-}}_{{\alpha}}+2 k'_{{\alpha}} ) \theta_{{\nu\lambda}}] g_{{\mu\kappa}} + [ - \theta_{{\alpha\kappa}}p^{-}_{{\nu}}+ \theta_{{\alpha\nu}}p^{-}_{{\kappa}} + \theta
_{{\alpha\kappa}}p^{+}_{{\nu}}- \theta_{{\alpha\nu}}p^{+}_{{\kappa}}
\\ \nonumber
& + &
 (  p^{-}_{{\alpha}}- p^{+}_{{\alpha}}) \theta_{{\nu\kappa}} ]g_{{\mu\lambda}} + [ 2 \theta_{{\alpha\kappa}}k'_{{\mu}}+ \theta_{{\mu
\alpha}}p^{-}_{{\kappa}}-\theta_{{\alpha\kappa}}p^{+}_{{\mu}}-\theta_{{\mu\alpha}}p^{+}_{{\kappa}}
\\ \nonumber
& + &
 (  p^{+}_{{\alpha}}+ p^{-}_{{\alpha}}+2 k'_{{\alpha}} ) \theta_
{{\mu\kappa}}] g_{{\nu\lambda}} +  [ -2 \theta_{{
\alpha\kappa}}k'_{{\lambda}}- \theta_{{\alpha\lambda}}{
p^{-}}_{{\kappa}}+\theta_{{\alpha\kappa}}p^{-}_{{\lambda}}+\theta_{{\alpha\lambda}}p^{+}_{{\kappa}}
\\ \nonumber
& + &
 (  p^{+}_{{\alpha}}+ p^{-}_{{\alpha}}+2 k'_{{\alpha}} ) \theta_
{{\kappa\lambda}} ] g_{{\mu\nu}} \} k_{{\alpha}}
 + \{  [  ( - \theta_{{\sigma\rho}}p^{+}_{{\sigma}}
+  \theta_{{\sigma\rho}}p^{-}_{{\sigma}} ) { g}_{{\mu\lambda}}
 - \theta_{{\mu\rho}}p^{-}_{{\lambda}}- \theta_{{\mu\lambda}}p^{-}_{{\rho}}
\\ \nonumber
& + &
2 \theta_{{\mu\rho}}k_{{\lambda}}- \theta_{{\rho\lambda}}p^{+}_{{\mu}}- \theta_{{\mu
\lambda}}p^{+}_{{\rho}}+2 \theta_{{\rho\lambda}}k_{{\mu}}] g_{{\nu\kappa}}
 +  (  \theta_{{\nu\rho}}{p^{+}}_{{\mu}}+ \theta_{{\mu\rho}}p^{-}_{{\nu}}+ \theta_{{\mu\nu}}p^{+}_{{\rho}}- \theta_{{\mu\rho}}p^{+}_{{\nu}}
\\ \nonumber
& - & 2 \theta_{{\nu\rho}}k_{{\mu}}+ \theta_{{\mu\nu}}p^{-}_{{\rho}}-
\theta_{{\sigma\rho}}p^{-}_{{\sigma}}g_{{\mu\nu}} ) g_{{\kappa\lambda}}
+ [  \theta_{{\sigma\rho}}p^{+}_{{\sigma}}g_{{\nu\lambda}}+ \theta_{{\nu\lambda}}p^{+}_{{\rho}}+ \theta_{{\nu\lambda}}p^{-}_\rho+2
\theta_{{\nu\rho}}k_{{\lambda}}- \theta_{{\nu\rho}} p^{-}_\lambda
\\ \nonumber
& + &
 ( - p^{-}_\nu + p^{+}_\nu )\theta_{{\rho\lambda}} ] g_{{\mu\kappa}}
 + [\theta_{{\nu\kappa}}p^{+}_{{\rho}}- \theta_{{\nu\kappa}}{ p^{-}}_{{\rho}}+ (  p^{-}_{{\kappa}}- p^{+}_{{\kappa}}) \theta_{{\nu\rho}}+ ( - p^{+}_{{\nu}}+ p^{-}_{{\nu}} ) \theta_{{\rho\kappa}} ] g_{{\mu\lambda}}
\\ \nonumber
 & + &
   (  \theta_{{\rho\kappa}}p^{+}_{{\mu}}- \theta_{{\mu\kappa}}p^{+}_{{\rho}}+ \theta_{{\mu\rho}}p^{+}_{{\kappa}}-2\theta_{{\mu\rho}}k_{{\kappa}}
   -2\theta_{{\rho\kappa}}{k}_{{\mu}}) g_{{\nu\lambda}}
  + ( - \theta_{{\rho\kappa}}p^{-}_{{\lambda}}+ \theta_{{\rho\lambda}}p^{-}_{{\kappa}}+2 \theta_{{\rho\kappa}}k_{{\lambda}}
\\ \nonumber
& - &
2 \theta_{{
\rho\lambda}}k_{{\kappa}}- \theta_{{\kappa\lambda}}{ p^{-}}_{{\rho}} ) g_{{\mu\nu}} \} k'_{{\rho}}+ [  ( \theta_{{\mu\rho}}k'_{{\lambda}}+\theta_{{\mu\rho}}k_{{\lambda}}+ \theta_{{\rho\lambda}}k_{{\mu}}+\theta_{{\rho\lambda}}k'_{{\mu}} )p^{+}_{{\rho}}
\\ \nonumber
& + &
 ( - \theta_{{\sigma\mu}}p^{-}_{{\sigma}}-\theta_{{\sigma
\mu}}p^{+}_{{\sigma}} ) k'_{{\lambda}} +  (
\theta_{{\rho\lambda}}k_{{\mu}}+ \theta_{{\rho\lambda}}{k'}_{{\mu}} ) p^{-}_{{\rho}} + ( - \theta_{{\sigma
\mu}}p^{-}_{{\sigma}}-\theta_{{\sigma\mu}}p^{+}_{{\sigma}}
 ) k_{{\lambda}} ] g_{{\nu\kappa}}
 \\ \nonumber
& + &
 [  ( \theta_{{\nu\rho}}k'_{{\mu}}-\theta_{{\mu\rho}
}k'_{{\nu}}+\theta_{{\nu\rho}}k_{{\mu}} ) { p^{+}}_{{\rho}}
 +  \theta_{{\nu\rho}}p^{-}_{{\rho}}k'_{{\mu}}+
 ( -\theta_{{\sigma\nu}}p^{-}_{{\sigma}}-\theta_{{\sigma\nu}
}p^{+}_{{\sigma}} ) k'_{{\mu}}-\theta_{{\sigma\nu}}{
 p^{+}}_{{\sigma}}k_{{\mu}}
 \\ \nonumber
& + &
( \theta_{{\sigma\mu}}{p^{+}}_{{\sigma}}+ \theta_{{\sigma\mu}}p^{-}_{{\sigma}} ) {k'}_{{\nu}} ] g_{{\kappa\lambda}} + [  ( -
 \theta_{{\rho\lambda}}k'_{{\nu}}-\theta_{{\nu\rho}}{ k'}
_{{\lambda}} ) p^{+}_{{\rho}}+ ( -\theta_{{\nu\rho}}{p^{-}}_{{\rho}}
 + \theta_{{\sigma\nu}}p^{-}_{{\sigma}}
 \\ \nonumber
 & + &
 \theta_{{
\sigma\nu}}p^{+}_{{\sigma}} ) k'_{{\lambda}} +  ( -
\theta_{{\nu\rho}}k_{{\lambda}}- \theta_{{\rho\lambda}}{
 k'}_{{\nu}} ) p^{-}_{{\rho}}+\theta_{{\sigma\nu}}{ p^{-}
}_{{\sigma}}k_{{\lambda}} ] g_{{\mu\kappa}}
  + [  ( -\theta_{{\nu\rho}}k_{{\kappa}}+ \theta_{{
\rho\kappa}}k'_{{\nu}} ) p^{+}_{{\rho}}
\\ \nonumber
& + &
 ( \theta_{{\sigma\nu}}p^{+}_{{\sigma}}+\theta_{{\nu\rho}}p^{-}_{{\rho}
}-\theta_{{\sigma\nu}}p^{-}_{{\sigma}} ) k_{{\kappa}}
 - \theta_{{\rho\kappa}}p^{-}_{{\rho}}k'_{{\nu}} ] g_{{\mu\lambda}}
  + [  ( -\theta_{{\mu\rho}}k_{{\kappa}}- \theta_{{\rho\kappa}}k'_{{\mu}}- \theta_{{\rho\kappa}}k_{{\mu}} ) p^{+}_{{\rho}}
\\ \nonumber
& + &
 ( \theta_{{\sigma\mu}}p^{+}_{{\sigma}}+ \theta_{{\sigma\mu}}p^{-}_{{\sigma}} ) k_{{\kappa}}- \theta_{{\rho\kappa}}{ p^{-}}_{{\rho}}k_{{\mu}} ] g_{{\nu\lambda}} + (\theta_{{\rho\kappa}}k_{{\lambda}} - \theta_{{\rho\lambda}
}k_{{\kappa}} ) g_{{\mu\nu}}p^{+}_{{\rho}}+ (  \theta_{{\rho\kappa}}p^{-}_{{\rho}}k'_{{\lambda}}
 \\ \nonumber
 & - &
 \theta_{{\rho\lambda}}p^{-}_{{\rho}}k_{{\kappa}}+
\theta_{{\rho\kappa}}p^{-}_{{\rho}}k_{{\lambda}} ) {
 g\_}_{{\mu\nu}} +  (  \theta_{{\mu\lambda}}p^{+}_{{\nu}}
+ \theta_{{\mu\nu}}p^{-}_{{\lambda}}+ \theta_{{\nu\lambda}}{
 p^{+}}_{{\mu}}+ \theta_{{\mu\lambda}}p^{-}_{{\nu}}
 - 2 \theta_{{\mu\nu}}k'_{{\lambda}}
\\ \nonumber
 & - & 2 \theta_{{\nu\lambda}}k'_{{\mu}}+2 \theta_{{\mu\lambda}}k'_{{\nu}} ) k_{{\kappa}} +  ( - \theta_{{\mu\nu}}p^{-}_{{\kappa}}- \theta_{{\mu
\nu}}p^{+}_{{\kappa}}- \theta_{{\nu\kappa}}p^{+}_{{\mu}}+2
\theta_{{\nu\kappa}}k_{{\mu}}- \theta_{{\mu\kappa}}{ p^{-}}_{{\nu}}
\\ \nonumber
& + &
 \theta_{{\mu\kappa}}p^{+}_{{\nu}} ) k'_{{\lambda}}
 +  ( - \theta_{{\mu\nu}}p^{-}_{{\kappa}} -  2 \theta_
{{\mu\kappa}}k'_{{\nu}} + \theta_{{\mu\nu}}p^{+}_{{\kappa}
}- \theta_{{\mu\kappa}}p^{-}_{{\nu}}- \theta_{{\mu\kappa}}{
 p^{+}}_{{\nu}}+ \theta_{{\nu\kappa}}p^{+}_{{\mu}}-2 \theta_{{
\nu\kappa}}k'_{{\mu}} ) k_{{\lambda}}
\\ \nonumber
& + &
 ( -
\theta_{{\kappa\lambda}}p^{+}_{{\nu}}- \theta_{{\nu\lambda}}{ p^{-}}_{{\kappa}}+ \theta_{{\nu\kappa}}p^{-}_{{\lambda}}+\theta_{{\kappa\lambda}}p^{-}_{{\nu}}
-  \theta_{{\nu\lambda}}{p^{+}}_{{\kappa}} ) k'_{{\mu}} +  ( -\theta_{{\nu
\kappa}}p^{-}_{{\lambda}}+ \theta_{{\nu\lambda}}p^{-}_{{\kappa}}
- \theta_{{\kappa\lambda}}p^{-}_{{\nu}}
 \\
 & - &
 2 \theta_{{\kappa\lambda}}k'_{{\nu}}- \theta_{{\kappa\lambda}}p^{+}_
{{\nu}}- \theta_{{\nu\lambda}}p^{+}_{{\kappa}} ) k_{{\mu}}
 +  (  \theta_{{\mu\lambda}}p^{+}_{{\kappa}}+ \theta
_{{\kappa\lambda}}p^{+}_{{\mu}}+ \theta_{{\mu\kappa}}p^{-}_{{\lambda}}+ \theta_{{\mu\lambda}}p^{-}_{{\kappa}} ) {k'}_{{\nu}} \bigl) \, .
\end{eqnarray}
\end{enumerate}
%
The symmetry properties of~\eqref{eq:SMWWZZ} -~\eqref{eq:nmWWZZ} are exactly the same as Eqs.~\eqref{eq:SMWWAA} -~\eqref{eq:nmWWAA}. The exchange of $Z^0$ bosons leaves the physical content of the diagram unchanged. The rules to be consistent with this requirement must be symmetric under substitutions $\nu \rightleftarrows \lambda$. Notice that, in~\eqref{eq:nmWWAA} momenta must be replaced with each other simultaneously as the indices are exchanged.
%
%
\item $W^{-} W^{+} Z \gamma$ coupling,
  \\
  \tikzset{
particle/.style={decorate, draw=black, decoration={snake=coil}},
}
\begin{tikzpicture}
[node distance=1.0cm and 1.5cm]
\coordinate[label=left:${A_\nu,q}$] (left up);
\coordinate[below right=of left up] (mid);
\coordinate[above right=of mid,label=right:${W^{+}_\kappa,{p}^+}$] (right up);
\coordinate[below left=of mid,label=left:${W^{-}_\mu,{p}^-}$] (left down);
\coordinate[below right=of mid,label=right:${Z_\lambda,{k}}$] (right down);
\coordinate[above=0.8cm of mid] (tem1);
\coordinate[left=0.8cm of tem1] (mom1i);
\coordinate[right=0.8cm of tem1] (mom2i);
\coordinate[above=0.35cm of mid] (tem2);
\coordinate[left=0.2cm of tem2] (mom1f);
\coordinate[right=0.2cm of tem2] (mom2f);
\coordinate[below=0.8cm of mid] (tem11);
\coordinate[left=0.8cm of tem11] (mom1ii);
\coordinate[right=0.8cm of tem11] (mom2ii);
\coordinate[below=0.35cm of mid] (tem22);
\coordinate[left=0.2cm of tem22] (mom1f1);
\coordinate[right=0.2cm of tem22] (mom2f2);
\draw[>=latex,->] (mom1ii) -- (mom1f1);
\draw[>=latex,->] (mom2ii) -- (mom2f2);
\draw[>=latex,->] (mom1i) -- (mom1f);
\draw[>=latex,->] (mom2i) -- (mom2f);
\draw[particle] (left up) -- (mid);
\draw[particle] (mid) -- (right up);
\draw[particle] (left down) -- (mid);
\draw[particle] (mid) -- (right down);
\end{tikzpicture}
\begin{enumerate}
\item Standard model
\begin{equation} \label{eq:SMWWZA}
- i e^2 \cot{\theta_{w}} \left( 2\, g_{\nu \lambda} g_{\mu \kappa} - g_{\nu \kappa} g_{\lambda \mu}
- g_{\mu \nu} g_{\kappa \lambda} \right) \, ,
\end{equation}
\item Minimal model
\begin{eqnarray} \label{eq:mWWZA} \nonumber
&\mathrm{Eq}&.~\eqref{eq:SMWWZA}
 +  \frac{1}{2} M^2_W e g \bigl[-6 \theta_{\kappa \lambda} g_{\nu \mu} - 2 \theta_{\kappa \mu} g_{\nu \lambda} - 3 \theta_{\nu \kappa} g_{\lambda \mu} + 2 \theta_{\mu \lambda} g_{\kappa \mu} + \theta_{\nu \mu} g_{\kappa \lambda}
 \\
 & + &
 (4 \theta_{\kappa \lambda} g_{\nu \mu} - 4 \theta_{\nu \kappa} g_{\lambda \mu}) \sin^2{\theta_{w}} \bigl] \, ,
\end{eqnarray}
\item Non-minimal model
\begin{eqnarray} \label{eq:nmWWZA} \nonumber
& \mathrm{Eq}&.~\eqref{eq:mWWZA} \, \, +  g' g^2 k_2 \bigl[ \cos^2{\theta_w} \bigl( ( \theta_{{\kappa \lambda}}g_{{\mu \nu}}- \theta_{{\mu \lambda}}g_{{\nu \kappa}}- \theta_{{\nu \lambda}}{\it
g}_{{\mu \kappa}}+ \theta_{{\mu \kappa}}g_{{\nu \lambda}}+
 \theta_{{\nu \kappa}}g_{{\mu \lambda}} ) {q_{{\alpha}}}^{2}
\\ \nonumber
& + &
 \{  ( - \theta_{{\alpha \kappa}}g_{{\nu \lambda}}+ \theta_{{\alpha \lambda}}g_{{\nu \kappa}} ) q_{{\mu}}+ [ 2\, \theta_{{\alpha \kappa}}q_{{\lambda}}
 + ( - \theta_{{\alpha \rho}}p^-_{{\rho}}-
\theta_{{\alpha \rho}}k_{{\rho}} ) g_{{\kappa
\lambda}}-2\, \theta_{{\alpha \lambda}}q_{{\kappa}}+ \theta_{{\alpha\kappa}}k_{{\lambda}}
\\ \nonumber
& + &
 (  p^-_{{\kappa}}-k_{{\kappa}} ) \theta_{{\alpha \lambda}}+ (  {
k}_{{\alpha}}+ p^-_{{\alpha}} ) \theta_{{\kappa \lambda}} ] g_{{\mu \nu}}+ [  \theta_{{\alpha \rho}}{p^-}_{{\rho}}g_{{\mu \lambda}}
 + ( - p^-_{{\mu}}+k_{{\mu}} ) \theta_{{\alpha \lambda}}
\\ \nonumber
& + &
 ( - { p^-}_{{\alpha}}- k_{{\alpha}} ) \theta_{{\mu \lambda}}+
\theta_{{\mu \alpha}}k_{{\lambda}} ] g_{{\nu\kappa}}+ [  \theta_{{\alpha \rho}}k_{{\rho}}g_{{\mu \kappa}}+ ( - k_{{\mu}}- p^-_{{\mu}}
) \theta_{{\alpha \kappa}}+ (  k_{{\alpha}}+ {p^-}_{{\alpha}} ) \theta_{{\mu \kappa}}
\\ \nonumber
& - &
\theta_{{\mu\alpha}}k_{{\kappa}}- \theta_{{\mu \alpha}}p^-_{{\kappa}} ] g_{{\nu \lambda}}+ (  \theta_{{\alpha \nu}}{ p^-}_{{\mu}}+ \theta_{{\alpha \nu}}k_{{\mu}}+ \theta_{{\mu \alpha}}p^-_{{\nu}}- \theta_{{\mu \nu}}p^-_{{\alpha}}+
 \theta_{{\mu \nu}}k_{{\alpha}} ) g_{{\kappa\lambda}}
\\ \nonumber
& + &
 (  \theta_{{\alpha \lambda}}g_{{\mu \kappa}}- \theta_{{\alpha \kappa}}g_{{\mu \lambda}} ) q_{{\nu}} + (  \theta_{{\nu \lambda}}k_{{\alpha}}-
\theta_{{\alpha \nu}}k_{{\lambda}} ) g_{{\mu\kappa}}+ ( - \theta_{{\nu\kappa}}p^-_{{\alpha}}-\theta_{{\alpha \nu}}p^-_{{\kappa}}+\theta_{{\alpha \kappa}}{ p^-}_{{\nu}} ) g_{{\mu \lambda}} \} q_{{\alpha}}
\\ \nonumber
& + &
 [ - \theta_{{\nu \kappa}}q_{{\lambda}}+ (  \theta_{{\rho \lambda}}k_{{\rho}}+ \theta_{{\rho\lambda}}p^-_{{\rho}} ) g_{{\nu \kappa}}
 + ( -2 \theta_{{\rho \kappa}}p^-_{{\rho}}+ \theta_{{\rho \kappa}}{ k}_{{\rho}} ) g_{{\nu \lambda}}+ (  \theta_{
{\nu \rho}}p^-_{{\rho}}+ \theta_{{\nu \rho}}k_{{\rho}}) g_{{\kappa \lambda}}
 \\ \nonumber
 & - &
  \theta_{{\kappa \lambda}}{ q}_{{\nu}}+ \theta_{{\nu \lambda}}q_{{\kappa}}- \theta
_{{\nu \kappa}}k_{{\lambda}}+ ( - k_{{\kappa}}-2 p^-_{{\kappa}} ) \theta_{{\nu \lambda}}
 - \theta_{{\kappa \lambda}}p^-_{{\nu}} ] q_{{\mu}}+ [
 (  \theta_{{\rho \kappa}}p^-_{{\rho}}- \theta_{{\rho\kappa}}k_{{\rho}} )q_{{\lambda}}
\\ \nonumber
& - &
 \theta_{{\rho \lambda}}q_{{\kappa}}p^-_{{\rho}} ] g_{{\mu \nu}}+ [  (  \theta_{{\mu \rho}}k_{{\rho}}+\theta_{{\mu \rho}}p^-_{{\rho}} ) g_{{\nu \kappa}}- \theta_{{\mu \kappa}}q_{{\nu}}- \theta_{{\nu \rho}}{ k}_{{\rho}}g_{{\mu \kappa}}- \theta_{{\mu \kappa}}p^-_{{\nu}}
 + (  k_{{\mu}}+ p^-_{{\mu}} ) \theta_{{\nu \kappa}}
\\ \nonumber
& + &
 \theta_{{\mu \nu}}p^-_{{\kappa}}- \theta_{{\mu\nu}}k_{{\kappa}} ] q_{{\lambda}}- \theta_{{\mu\rho}}q_{{\kappa}}p^-_{{\rho}}g_{{\nu \lambda}}-2
\ \theta_{{\mu \rho}}q_{{\nu}}k_{{\rho}}g_{{\kappa \lambda}}+ ( - \theta_{{\mu \kappa}}k_{{\lambda}}+ \theta_{{\mu \lambda}}k_{{\kappa}}- \theta_{{\rho \lambda}}k_{{\rho}}g_{{\mu \kappa}}
\\ \nonumber
& - &
 \theta_{{\kappa \lambda}}k_{{\mu}}+ \theta_{{\mu \lambda}}q_{{\kappa}}
 ) q_{{\nu}}- \theta_{{\nu \rho}}q_{{\kappa}}{ p^-}_{{\rho}}g_{{\mu \lambda}}+ (  \theta_{{\mu\lambda}}p^-_{{\nu}}+ \theta_{{\nu \lambda}}p^-_{{\mu}}
 ) q_{{\kappa}} \bigl)
\\ \nonumber
& + & \sin^2{\theta_{w}}
 \bigl( \{  [  ( - \theta_{{\sigma \eta}}p^-_{{\sigma}}+ \theta_{{\sigma\eta}}q_{{\sigma}} ) g_{{\kappa \lambda}}+ ( - p^+_{{\lambda}}+ q_{{\lambda}} ) \theta_{{\eta \kappa}}+ (  p^+_{{\kappa}}- p^-_{{\kappa}} ) \theta_{{\eta \lambda}}
\\ \nonumber
& + &
 ( k_{{\eta}} )\theta_{{\kappa \lambda}} ] g_{{\mu \nu}}+ [( - \theta_{{\sigma \eta}}p^+_{{\sigma}}+ \theta_{{\sigma \eta}}p^-_{{\sigma}} ) g_{{\mu \lambda}}+
 (  p^-_{{\mu}}- q_{{\mu}} ) \theta_{{\eta\lambda}}
\\ \nonumber
& + &
 (  p^+_{{\lambda}}- q_{{\lambda}} ) \theta_{{\mu \eta}}+ ( -k_{{\eta}} ) \theta_{{\mu \lambda}} ] {g}_{{\nu \kappa}} + [  (  \theta_{{\sigma \eta}}{
p^+}_{{\sigma}}- \theta_{{\sigma \eta}}q_{{\sigma}} ) {g}_{{\mu \kappa}}+ (  p^-_{{\mu}}- q_{{\mu}} ) \theta_{{\eta \kappa}}
\\ \nonumber
& + &
 (  p^-_{{\kappa}}- { p^+}_{{\kappa}} ) \theta_{{\mu \eta}}+ ( k_{{\eta}} ) \theta_{{\mu\kappa}} ] g_{{\nu \lambda}}+ [  (  p^+_{{\lambda}}- q_{{\lambda}} ) \theta_{{\eta\nu}}
 + (  q_{{\nu}}- p^+_{{\nu}} ) \theta_{{\eta\lambda}}
\\ \nonumber
& + &
 (  p^+_{{\eta}}- q_{{\eta}} )\theta_{{\nu \lambda}} ] g_{{\mu \kappa}}+ [(  p^-_{{\kappa}}- p^+_{{\kappa}} ) \theta_{
{\eta \nu}}+ (  p^+_{{\nu}}- p^-_{{\nu}} )\theta_{{\eta \kappa}}+ ( - p^+_{{\eta}}+ p^-_{{\eta}} ) \theta_{{\nu \kappa}} ] g_{{\mu \lambda}}
\\ \nonumber
& + &
 [  ( - p^-_{{\mu}}+ q_{{\mu}} )\theta_{{\eta \nu}}+ ( - p^-_{{\nu}}+ q_{{\nu}} ) \theta_{{\mu \eta}}+ ( - q_{{\eta}}+ p^-
_{{\eta}} ) \theta_{{\mu \nu}} ] g_{{\kappa\lambda}} \} k_{{\eta}}+ [  (  \theta_{{\sigma\lambda}}p^+_{{\sigma}}+ \theta_{{\sigma\lambda}}q_{{\sigma}}
\\ \nonumber
& + &
 \theta_{{\sigma \lambda}}p^-_{{\sigma}} ) {k}_{{\kappa}}
 + ( \theta_{{\sigma \kappa}}k_{{\sigma}} ) k_{{\lambda}} ] {g}_{{\mu \nu}}+ [  ( \theta_{{\sigma \lambda}}k_{{\sigma}} ) k_{{\mu}}
  - \theta_{{\sigma \mu}}k_{{\sigma}} k_{{\lambda}} ] g_{{\nu \kappa}}
 \\ \nonumber
& + & [  ( - \theta_{{\sigma \kappa}}k_{{\sigma}} ) k_{{\mu}}+ (  \theta_{{\sigma \mu}}k_{{\sigma}}) k_ {{\kappa}} ] g_{{\nu \lambda}}
+ [  (  \theta_{{\sigma \nu}}p^-_{{\sigma}}- \theta_{{\sigma \nu}}q_{{\sigma}} ) g_{{\kappa \lambda}}+ (  p^+_{{\lambda}}- q_{{\lambda}} ) \theta_{{\nu \kappa}}
 + (  p^-_{{\kappa}}- p^+_{{\kappa}} ) \theta_{{\nu \lambda}}
\\ \nonumber
& - &
 ( k_{{\nu}} ) \theta_{{\kappa \lambda}} ] k_{{\mu}}+ [  ( - \theta_{{\sigma \nu}}p^-_{{\sigma}}+
\theta_{{\sigma \nu}}p^+_{{\sigma}} ) g_{{\mu\lambda}}+ ( - p^+_{{\lambda}}+ q_{{\lambda}}) \theta_{{\mu \nu}}
+ ( k_{{\nu}} ) \theta_{{\mu \lambda}}
+( -p^-_{{\mu}}+ q_{{\mu}} ) \theta_{{\nu \lambda}} ] k_{{\kappa}}
\\
& + & [  ( - \theta_{{\sigma \nu}}
p^+_{{\sigma}}+ \theta_{{\sigma \nu}}q_{{\sigma}}) g_{{\mu \kappa}}
 +
  (  p^+_{{\kappa}}- { p^-}_{{\kappa}} ) \theta_{{\mu \nu}}- (  k_{{\nu}} ) \theta_{{\mu\kappa}}+ ( - p^-_{{\mu}}+ q_{{\mu}} )\theta_{{\nu \kappa}} ] k_{{\lambda}} \bigl) \bigl] \,.
\end{eqnarray}
\end{enumerate}
The rule~\eqref{eq:SMWWZA} is the standard model vertex function for $W^-W^+ Z \gamma$ coupling. In the present case there is no explicit exchange symmetry. Equation~\eqref{eq:mWWZA} represents the $\theta$-expanded rule for the minimal noncommutative model. Its $\theta$-dependent part is obtained from the Higgs induced interactions. Equation~\eqref{eq:nmWWZA} is the Feynman rule for non-minimal extended model. The momentum dependent part of~\eqref{eq:nmWWZA} is derived from the gauge action~\eqref{eq:nmrulesphysical}.
%
\item $\gamma g g g$ coupling,
 \\
 \tikzset{
particle/.style={decorate, draw=black, decoration={snake=coil}},
gluon/.style={decorate, draw=black, decoration={coil,aspect=1.2}},
}
\begin{tikzpicture}
[node distance=1.0cm and 1.5cm]
\coordinate[label=left:${G_\nu^{a} ,p}$] (left up);
\coordinate[below right=of left up] (mid);
\coordinate[above right=of mid,label=right:${G_\kappa^{b} ,p'}$] (right up);
\coordinate[below left=of mid,label=left:${A_\mu, q}$] (left down);
\coordinate[below right=of mid,label=right:${G_\lambda^{c},p''}$] (right down);
\coordinate[above=0.8cm of mid] (tem1);
\coordinate[left=0.8cm of tem1] (mom1i);
\coordinate[right=0.8cm of tem1] (mom2i);
\coordinate[above=0.35cm of mid] (tem2);
\coordinate[left=0.2cm of tem2] (mom1f);
\coordinate[right=0.2cm of tem2] (mom2f);
\coordinate[below=0.8cm of mid] (tem11);
\coordinate[left=0.8cm of tem11] (mom1ii);
\coordinate[right=0.8cm of tem11] (mom2ii);
\coordinate[below=0.35cm of mid] (tem22);
\coordinate[left=0.2cm of tem22] (mom1f1);
\coordinate[right=0.2cm of tem22] (mom2f2);
\draw[>=latex,->] (mom1ii) -- (mom1f1);
\draw[>=latex,->] (mom2ii) -- (mom2f2);
\draw[>=latex,->] (mom1i) -- (mom1f);
\draw[>=latex,->] (mom2i) -- (mom2f);
\draw[gluon] (left up) -- (mid);
\draw[gluon] (mid) -- (right up);
\draw[particle] (left down) -- (mid);
\draw[gluon] (mid) -- (right down);
\end{tikzpicture}
\begin{enumerate}
\item Non-minimal model
\begin{eqnarray} \label{eq:AGGG} \nonumber
& g'& g^3_{\mathrm{s}}\cos{\theta_{w}} f^{{b a c}} \bigl( \{  [ 2 \theta_{{\sigma \rho}}{ p'}_{{\sigma}}{g}_{{\kappa \lambda}}-2 \theta_{{\sigma \rho}}{ p''}_{{\sigma}}g_{{\kappa \lambda}}+ ( 2 { p'}_{{\lambda}}-2 {p}_{{\lambda}} ) \theta_{{\rho \kappa}}
 \\ \nonumber
& + &
( -2 { p''}_{{\kappa}}+2 { p}_{{\kappa}} ) \theta_{{\rho \lambda}}
 ] g_{{\nu \mu}} +  [ -2 \theta_{{\sigma \rho}}{p}_{{\sigma}}g_{{\nu\lambda}}+ 2 \theta_{{\sigma \rho}}{p''}_{{\sigma}}g_{{\nu \lambda}}+ ( -2 { p'}_{{\lambda}}+2 {p}_{{\lambda}} ) \theta_{{\nu \rho}}
\\ \nonumber
& + &
 ( 2 {p''}_{{\nu}}-2 { p'}_{{\nu}} ) \theta_{{\rho \lambda}}
 ] g_{{\kappa \mu}}+ [ 2 \theta_{{\sigma \rho}}{ p}_{{\sigma}}g_{{\nu \kappa}}-2 \theta_{{\sigma \rho}}{ p'}_{{\sigma}}g_{{\nu \kappa}}
 + ( 2 { p''}_{{\kappa}}-2 { p}_{{\kappa}} ) \theta_{{\nu \rho}}
\\ \nonumber
& + &
 ( 2
 { p''}_{{\nu}}-2 { p'}_{{\nu}} ) \theta_{{\rho \kappa}} ] g_{{\lambda\mu}} +  (  ( -2 { p'}_{{\lambda}}+2 { p}_{{\lambda}} ) \theta_{{\rho\mu}}+ (-2 { p}_{{\mu}}+2 { p'}_{{\mu}} ) \theta_{{\rho \lambda}} ) g_{{\nu\kappa}}
\\ \nonumber
& + & [  ( 2 { p''}_{{\kappa}}-2 { p}_{{\kappa}} ) \theta_{{\rho\mu}}+ ( -2{ p''}_{{\mu}}+2 { p}_{{\mu}} ) \theta_{{\rho \kappa}}] g_{{\nu \lambda}} +  [  ( -2 { p''}_{{\mu}}+2 { p'}_{{\mu}} ) \theta_{{\nu\rho}}
\\ \nonumber
& + & ( 2 { p'}_{{\nu}}-2 { p''}_{{\nu}} ) \theta_{{\rho \mu}} ] {g}_{{\kappa \lambda}} \} { q}_{{\rho}} +  [  -2 \theta_{{\sigma \lambda}}{ q}_{{
\sigma}}  { q}_{{\kappa}} + 2 \theta_{{\sigma\kappa}}{ q}_{{\sigma}} {q}_{{\lambda}}
+ 2 \theta_{{\kappa\lambda}}{ q}_{{\sigma}} { q}_{{\sigma}} ] {g}_{{\nu \mu}}
 +  [ 2 \theta_{{\sigma \lambda}}{ q}_{{\sigma}} { q}_{{\nu}}
\\ \nonumber
& - & 2 \theta_{{\sigma \nu}}{ q}_{{\sigma}} { q}_{{\lambda}} - 2 \theta_{{\nu \lambda}}{ q}_{{\sigma}} { q}_{{\sigma}} ] {g}_{{\kappa\mu}}
+ [  -2\theta_{{\sigma \kappa}}{ q}_{{\sigma}} { q}_{{\nu}}+ 2 \theta_{{\sigma \nu}}{ q}_{{\sigma}} { q}_{{\kappa}}
+  2 \theta_{{\nu \kappa}}{ q }_{{\sigma}} {q}_{{\sigma}}] g_{{\lambda \mu}} +  [ -2 \theta_{{\sigma \mu}}{p'}_{{\sigma}}g_{{\kappa \lambda}}
 \\ \nonumber
 & + &
 2 \theta_{{\sigma \mu}}{ p''}_{{\sigma}}g_{{\kappa \lambda}}+ ( 2 { p'}_{
{\lambda}}-2 { p}_{{\lambda}} ) \theta_{{\kappa \mu}} -2 { q}_{{\mu}} \theta_{{\kappa \lambda}} + ( -2 { p''}_{{\kappa}}+2{ p}_{{\kappa}} ) \theta_{{\lambda \mu}} ] {q}_{{\nu}}
 \\ \nonumber
 & + & [2 \theta_{{\sigma \mu}}{ p}_{{\sigma}}g_{{\nu \lambda}}-2 \theta_{{\sigma \mu}}{ p''}_{{\sigma}}g_{{\nu \lambda}}+ ( 2 {p'}_{{\lambda}}-2 { p}_{{\lambda}}) \theta_{{\nu \mu}} + 2 { q}_{{\mu}} \theta_{{\nu \lambda}}+ ( 2 { p''}_{{\nu}}-2 { p'}_{{\nu}} ) \theta_{{\lambda \mu}}
 ] { q}_{{\kappa}}
\\ \nonumber
 & + &  [ -2 \theta_{{\sigma \mu}}{ p}
_{{\sigma}}g_{{\nu \kappa}}+2 \theta_{{\sigma \mu}}{ p'}_{{\sigma}}g_{{\nu \kappa}}
+ ( -2 { p''}_{{\kappa}}+2 { p}_{{\kappa}} ) \theta_{{\nu \mu}}
  - 2 {q}_{{\mu}} \theta_{{\nu
\kappa}}+ ( 2 { p''}_{{\nu}}-2 { p'}_{{\nu}} )
\theta_{{\kappa \mu}} ] { q}_{{\lambda}}
\\
& + & [  ( 2\theta_{{\kappa \mu}}g_{{\nu \lambda}}-2 \theta_{{\lambda \mu}}g_{{\nu \kappa}} ) { p}_{{\sigma}}+ ( 2
\theta_{{\lambda \mu}}g_{{\nu \kappa}}-2 \theta_{{\nu \mu}}{g}_{{\kappa \lambda}} ) { p'}_{{\sigma}}
 -( 2\theta_{{\kappa \mu}}g_{{\nu \lambda}}-2 \theta_{{\nu \mu}}{g}_{{\kappa \lambda}} ) { p''}_{{\sigma}} ] {q}_{{\sigma}} \bigl) \, . \, \,
\end{eqnarray}
\end{enumerate}
%
The $\gamma ggg$ coupling is forbidden in the standard model. The rule~\eqref{eq:AGGG} is derived from~\eqref{eq:nmrulesqcd} and is allowed only in the non-minimal noncommutative model. Because the exchange of gluons leaves the diagram topologically invariant, the rule must be symmetric under simultaneous substitutions of $\nu \rightleftarrows \kappa$, $a \rightleftarrows b$, $p \rightleftarrows p'$ and independently under $\nu \rightleftarrows \lambda$, $a \rightleftarrows c$, $p \rightleftarrows p''$ as well as under $\kappa \rightleftarrows \lambda$, $b \rightleftarrows c$, $p' \rightleftarrows p''$. Also, the cyclic symmetry $\nu \rightleftarrows \kappa \rightleftarrows \lambda$, $a \rightleftarrows b \rightleftarrows c$, $p \rightleftarrows p' \rightleftarrows p''$ must be satisfied.
%
\item $Z g g g$ coupling,
\\
 \tikzset{
particle/.style={decorate, draw=black, decoration={snake=coil}},
gluon/.style={decorate, draw=black, decoration={coil,aspect=1.2}},
}
\begin{tikzpicture}
[node distance=1.0cm and 1.5cm]
\coordinate[label=left:${G_\nu^{a} ,p}$] (left up);
\coordinate[below right=of left up] (mid);
\coordinate[above right=of mid,label=right:${G_\kappa^{b} ,p'}$] (right up);
\coordinate[below left=of mid,label=left:${Z_\mu, k}$] (left down);
\coordinate[below right=of mid,label=right:${G_\lambda^{c},p''}$] (right down);
\coordinate[above=0.8cm of mid] (tem1);
\coordinate[left=0.8cm of tem1] (mom1i);
\coordinate[right=0.8cm of tem1] (mom2i);
\coordinate[above=0.35cm of mid] (tem2);
\coordinate[left=0.2cm of tem2] (mom1f);
\coordinate[right=0.2cm of tem2] (mom2f);
\coordinate[below=0.8cm of mid] (tem11);
\coordinate[left=0.8cm of tem11] (mom1ii);
\coordinate[right=0.8cm of tem11] (mom2ii);
\coordinate[below=0.35cm of mid] (tem22);
\coordinate[left=0.2cm of tem22] (mom1f1);
\coordinate[right=0.2cm of tem22] (mom2f2);
\draw[>=latex,->] (mom1ii) -- (mom1f1);
\draw[>=latex,->] (mom2ii) -- (mom2f2);
\draw[>=latex,->] (mom1i) -- (mom1f);
\draw[>=latex,->] (mom2i) -- (mom2f);
\draw[gluon] (left up) -- (mid);
\draw[gluon] (mid) -- (right up);
\draw[particle] (left down) -- (mid);
\draw[gluon] (mid) -- (right down);
\end{tikzpicture}
\begin{enumerate}
\item Non-minimal model
\begin{eqnarray} \label{eq:ZGGG} \nonumber
& g'& g^3_{\mathrm{s}}\sin{\theta_{w}} f^{{a b c}} \bigl( \{  [ 2 \theta_{{\sigma \rho}}{ p'}_{{\sigma}}{g}_{{\kappa \lambda}}-2 \theta_{{\sigma \rho}}{ p''}_{{\sigma}}g_{{\kappa \lambda}}+ ( 2 { p'}_{{\lambda}}-2 {p}_{{\lambda}} ) \theta_{{\rho \kappa}}
 \\ \nonumber
& + & ( -2 { p''}_{{\kappa}}+2 { p}_{{\kappa}} ) \theta_{{\rho \lambda}}
 ] g_{{\nu \mu}}
[ -2 \theta_{{\sigma \rho}}{p}_{{\sigma}}g_{{\nu\lambda}}+ 2 \theta_{{\sigma \rho}}{p''}_{{\sigma}}g_{{\nu \lambda}}+ ( -2 { p'}_{{\lambda}}+2 {p}_{{\lambda}} ) \theta_{{\nu \rho}}
\\ \nonumber
& + & ( 2 {p''}_{{\nu}}-2 { p'}_{{\nu}} ) \theta_{{\rho \lambda}}
 ] g_{{\kappa \mu}}
 [ 2 \theta_{{\sigma \rho}}{ p}_{{\sigma}}g_{{\nu \kappa}}-2 \theta_{{\sigma \rho}}{ p'}_{{\sigma}}g_{{\nu \kappa}}
 + ( 2 { p''}_{{\kappa}}-2 { p}_{{\kappa}} ) \theta_{{\nu \rho}}
\\ \nonumber
& + & ( 2
 { p''}_{{\nu}}-2 { p'}_{{\nu}} ) \theta_{{\rho \kappa}} ] g_{{\lambda\mu}}
(  ( -2 { p'}_{{\lambda}}+2 { p}_{{\lambda}} ) \theta_{{\rho\mu}}+ (-2 { p}_{{\mu}}+2 { p'}_{{\mu}} ) \theta_{{\rho \lambda}} ) g_{{\nu\kappa}}
\\ \nonumber
& + & [  ( 2 { p''}_{{\kappa}}-2 { p}_{{\kappa}} ) \theta_{{\rho\mu}}+ ( -2{ p''}_{{\mu}}+2 { p}_{{\mu}} ) \theta_{{\rho \kappa}}] g_{{\nu \lambda}} +  [  ( -2 { p''}_{{\mu}}+2 { p'}_{{\mu}} ) \theta_{{\nu\rho}}
\\ \nonumber
& + & ( 2 { p'}_{{\nu}}-2 { p''}_{{\nu}} ) \theta_{{\rho \mu}} ] {g}_{{\kappa \lambda}} \} { k}_{{\rho}} +  [  -2 \theta_{{\sigma \lambda}}{ k}_{{
\sigma}}  { k}_{{\kappa}} + 2 \theta_{{\sigma\kappa}}{ k}_{{\sigma}} {k}_{{\lambda}}
+ 2 \theta_{{\kappa\lambda}}{ k}_{{\sigma}} { k}_{{\sigma}} ] {g}_{{\nu \mu}}
 +  [ 2 \theta_{{\sigma \lambda}}{ k}_{{\sigma}} { k}_{{\nu}}
\\ \nonumber
& - & 2 \theta_{{\sigma \nu}}{ k}_{{\sigma}} { k}_{{\lambda}} - 2 \theta_{{\nu \lambda}}{ k}_{{\sigma}} { k}_{{\sigma}} ] {g}_{{\kappa\mu}}
+ [  -2\theta_{{\sigma \kappa}}{ k}_{{\sigma}} { k}_{{\nu}}+ 2 \theta_{{\sigma \nu}}{ k}_{{\sigma}} { k}_{{\kappa}}
+  2 \theta_{{\nu \kappa}}{ k }_{{\sigma}} {k}_{{\sigma}}] g_{{\lambda \mu}} +  [ -2 \theta_{{\sigma \mu}}{p'}_{{\sigma}}g_{{\kappa \lambda}}
 \\ \nonumber
 & + &
 2 \theta_{{\sigma \mu}}{ p''}_{{\sigma}}g_{{\kappa \lambda}}+ ( 2 { p'}_{
{\lambda}}-2 { p}_{{\lambda}} ) \theta_{{\kappa \mu}} -2 { k}_{{\mu}} \theta_{{\kappa \lambda}} + ( -2 { p''}_{{\kappa}}+2{ p}_{{\kappa}} ) \theta_{{\lambda \mu}} ] {k}_{{\nu}}
 \\ \nonumber
 & + & [2 \theta_{{\sigma \mu}}{ p}_{{\sigma}}g_{{\nu \lambda}}-2 \theta_{{\sigma \mu}}{ p''}_{{\sigma}}g_{{\nu \lambda}}+ ( 2 {p'}_{{\lambda}}-2 { p}_{{\lambda}}) \theta_{{\nu \mu}} + 2 { k}_{{\mu}} \theta_{{\nu \lambda}}+ ( 2 { p''}_{{\nu}}-2 { p'}_{{\nu}} ) \theta_{{\lambda \mu}}
 ] { k}_{{\kappa}}
\\ \nonumber
 & + &  [ -2 \theta_{{\sigma \mu}}{ p}
_{{\sigma}}g_{{\nu \kappa}}+2 \theta_{{\sigma \mu}}{ p'}_{{\sigma}}g_{{\nu \kappa}}
+ ( -2 { p''}_{{\kappa}}+2 { p}_{{\kappa}} ) \theta_{{\nu \mu}}
  - 2 {k}_{{\mu}} \theta_{{\nu
\kappa}}+ ( 2 { p''}_{{\nu}}-2 { p'}_{{\nu}} )
\theta_{{\kappa \mu}} ] { k}_{{\lambda}}
\\
& + & [  ( 2\theta_{{\kappa \mu}}g_{{\nu \lambda}}-2 \theta_{{\lambda \mu}}g_{{\nu \kappa}} ) { p}_{{\sigma}}+ ( 2
\theta_{{\lambda \mu}}g_{{\nu \kappa}}-2 \theta_{{\nu \mu}}{g}_{{\kappa \lambda}} ) { p'}_{{\sigma}}
 -( 2\theta_{{\kappa \mu}}g_{{\nu \lambda}}-2 \theta_{{\nu \mu}}{g}_{{\kappa \lambda}} ) { p''}_{{\sigma}} ] {k}_{{\sigma}} \bigl) \, . \,\,
\end{eqnarray}
\end{enumerate}
%
%
The $Z ggg$ coupling is also forbidden in the standard model at tree level. This vertex function is derived from~\eqref{eq:nmrulesqcd2} and is allowed only in the non-minimal model. The symmetry properties of~\eqref{eq:ZGGG} are the same as that of the rule~\eqref{eq:AGGG}.
\end{enumerate}
\section{Discussion on phenomenological perspectives of the model}\label{discuss}
\begin{figure}[H]
\captionwidth 2.0in
\begin{minipage}[H]{2.2 in}
\includegraphics[width=2.0 in]{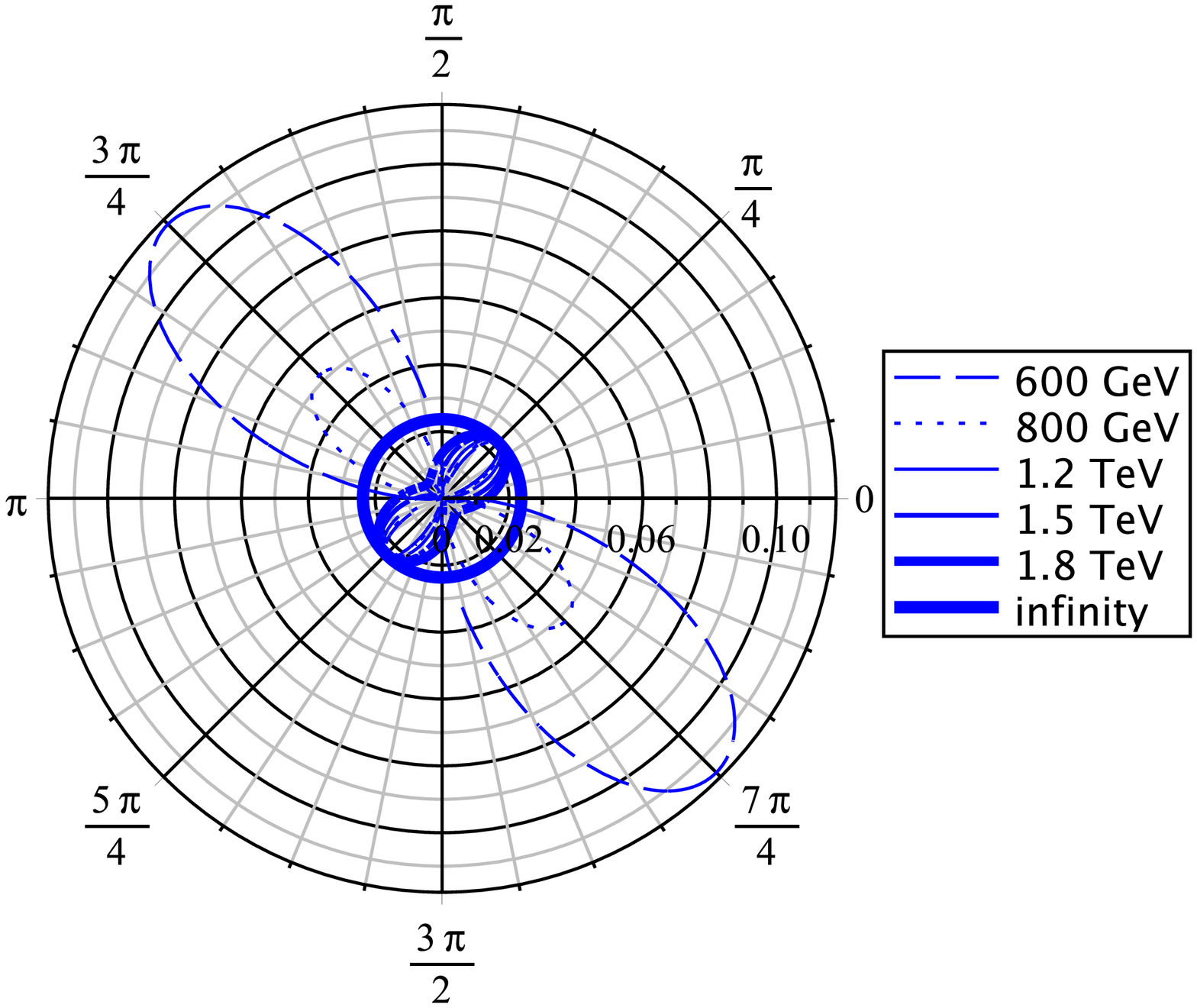}
\caption{Differential distributions $\frac{d\sigma}{d\phi}$ at ${\theta=\frac{\pi}{2}}$ in SM  for $\sqrt{s}=1.0, 1.5, 2.0$ TeV.}
\label{fig:figure1}
\end{minipage}
\begin{minipage}[H]{2.2 in}
\includegraphics[width=2.0 in]{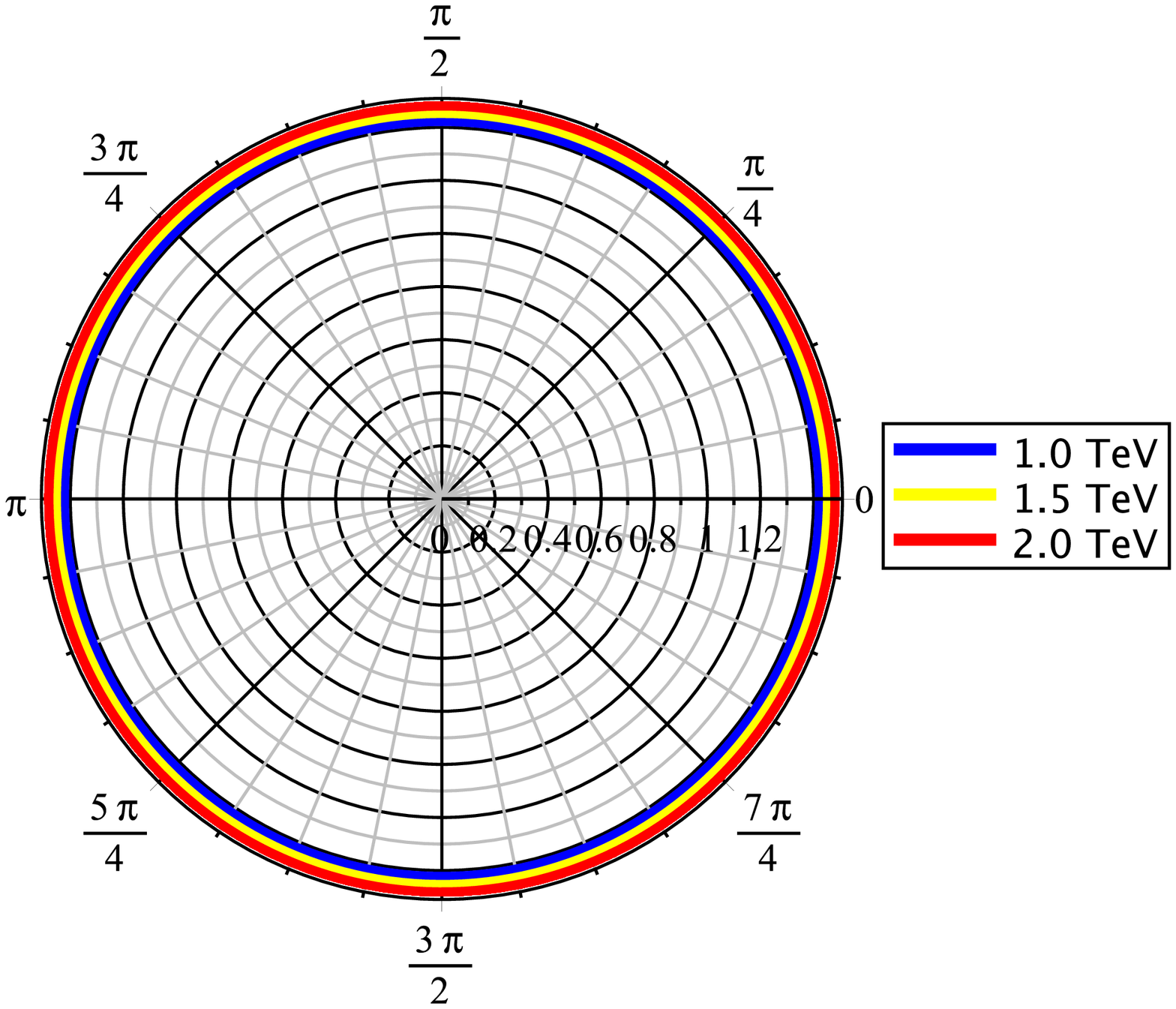}
\caption{The $\theta-$integrated $\frac{d\sigma}{d\phi}$ distributions in SM for $\sqrt{s}=1.0, 1.5, 2.0$ TeV.}
\label{fig:figure2}
\end{minipage}
\begin{minipage}[H]{2.2 in}
\includegraphics[width=2.0 in]{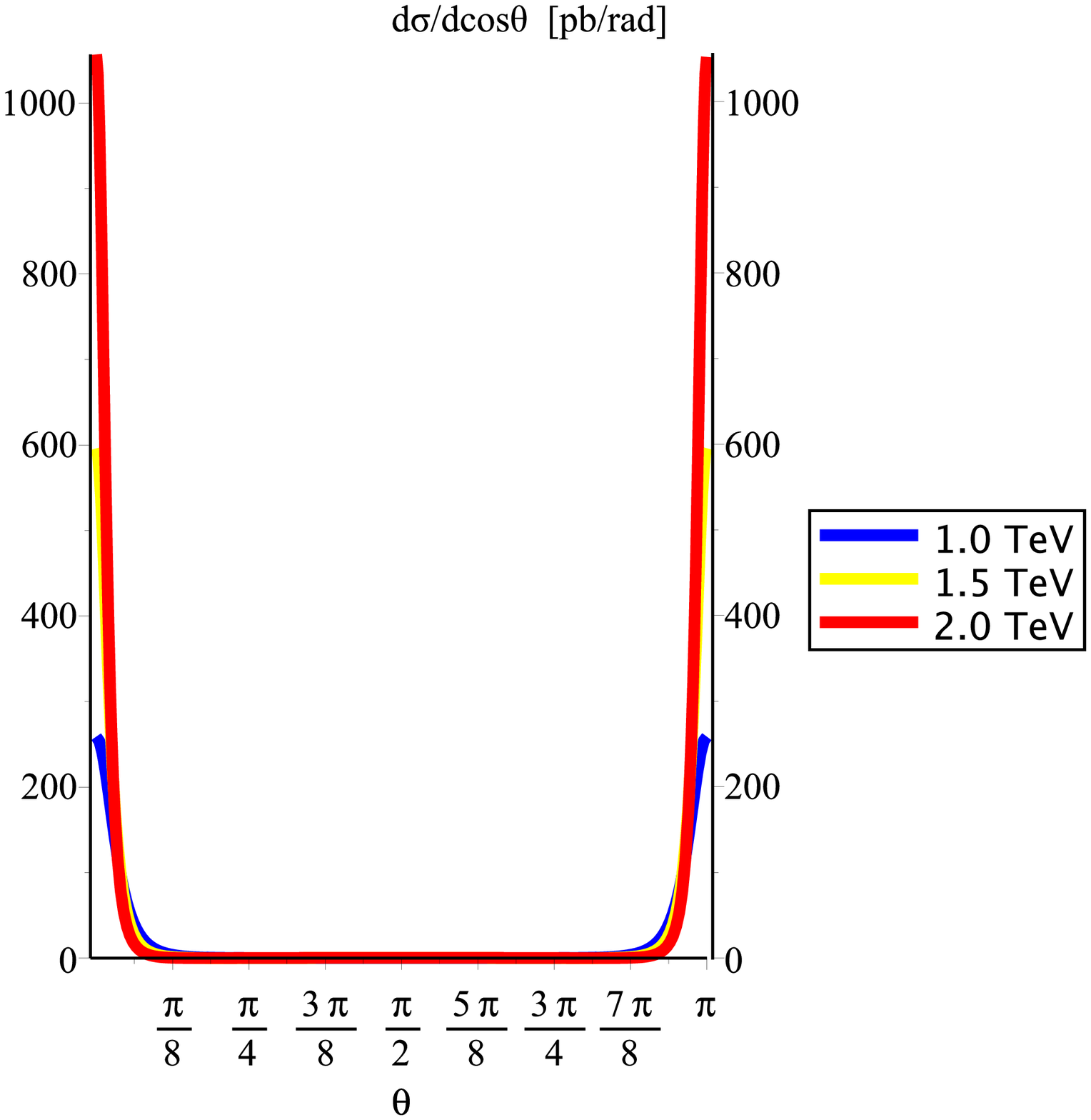}
\caption{The $\phi-$integrated $\frac{d\sigma}{d\cos{\theta}}$ distributions in SM for $\sqrt{s}=1.0, 1.5, 2.0$ TeV.}
\label{fig:figure3}
\end{minipage}
\end{figure}

To give an intuitive understanding of the model and in particular, the Feynman rules developed in the previous section, let us consider the $W^-W^+ \rightarrow Z Z$ scattering. In the context of the standard model and at tree level, the scattering amplitude of process is the sum over amplitudes of diagrams~\ref{contact} -~\ref{schannelhiggs}, in Appendix~\ref{diag}. Using the analysis of relevant diagrams the scattering cross section is estimated approximately to be around 68 pb ($10^{-12}$ barn). The amplitude for the contact interaction grows rapidly as the centre of mass (c.m.) energy $\sqrt{s}$ increases. The amplitudes of $t$-channel and the exchange diagrams give rise respectively to forward and backward scattering. On the other hand, the Higgs mediated diagram effectively tames the amplitude of the contact interaction and there is a strong cancellation in the high energy behaviour of individual amplitudes. The total cross section ultimately reaches to a nearly constant value at large c.m. energies. It is well known that, only the scattering of longitudinally polarized bosons is responsible for the leading behaviour of scattering amplitudes at high energy limit. Then, let us define the kinematics of scattering as
\begin{subequations}\label{smgaugeaction} 
\begin{eqnarray}
p^\pm & = & \left( E, \, 0 \,, \, 0 \,, \, \pm p\right), \,
\\
k^\pm & = & \left( E, \, 0, \, \pm p \sin{\theta}, \, \pm p \cos{\theta} \right), \,
\\
\varepsilon^\pm(p) & = &  \left( \frac{p}{M_W}, \, 0 \,, \, 0 \,, \, \pm \frac{E}{M_W}\right),
\\
\varepsilon^\pm(k) & = &  \left( \frac{p}{M_Z}, \, 0\, , \, \pm \frac{E \sin{\theta}}{M_Z} \, , \, \pm \frac{E \cos{\theta}}{M_Z} \right).
\end{eqnarray}
\end{subequations}
Here, $\theta$ and $\phi$ are respectively the polar and azimuthal angle. The momenta of incoming $W^\pm$  and outgoing $Z^0$ bosons in the c.m. reference frame are respectively denoted by $p^{\pm}$  and $k^{\pm}$. Also, $\varepsilon^\pm(p)$ and $\varepsilon^\pm(k)$ are used for polarization vectors of corresponding bosons. The general features of the scattering can be understood from Figs.~\ref{fig:figure1} -~\ref{fig:figure3}. In the standard model, the azimuthal distribution of differential cross section, i.e., $\frac{d\sigma}{d\phi}$, is expected to be flat. The $\frac{d\sigma}{d\phi}$ distributions at $\theta=\frac{\pi}{2}$ have been shown in Fig.~\ref{fig:figure1}, for $\sqrt{s}=$ 1.0, 1.5, and 2.0 TeV. Figure~\ref{fig:figure2}, displays the $\theta-$integrated $\frac{d\sigma}{d\phi}$ distributions at same c.m. energies. On the other hand, because of the $t$ and $u-$channel diagrams the $\phi-$integrated cross sections $\frac{d\sigma}{d\cos \theta}$ are expected to be very forward-backward distributions. The $\frac{d\sigma}{d\cos \theta}$ distributions have been shown in Fig.~\ref{fig:figure3}. For numerical evaluations some approximations were made. We neglected the decay width of intermediate bosons and assumed they were being nearly stable particles. Also, the integrations on $\theta$ and $\phi$ were performed approximately because the subsequent similar integrations on $\frac{d\sigma_{\mathrm{NC}}}{d\Omega}$ ($d\Omega = \sin{\theta} d\theta d\phi$) could not be evaluated exactly in a reasonable computation time. We preferred to use the same level of accuracy for all of numerical evaluations from the beginning. The masses of weak bosons are $M_W = 80.38$, $M_Z = 91.18$ and the Higgs mass is $M_H = 125.0$ GeV based on last issue of Particle Data Group~\cite{pdg}. Furthermore, the parameter $k_2$ (see the rule~\eqref{eq:nmWWZZ}) is assumed to be 0.4~\cite{ana2}. In our evaluations the total cross section approaches to 28 pb at $\sqrt{s}= 1.5$ TeV. This is about 12.5\% smaller than the exact value 32 pb~\cite{ana3}.

Now, we consider the process in the context of noncommutative standard model. The usual parametrization for deformation quantity is $\theta^{\mu\nu} = {c^{\mu\nu}}/{\Lambda^2_{\mathrm{NC}}}$ where, $c^{\mu\nu}$ is a dimensionless matrix of order unity and $\Lambda_{\mathrm{NC}}$ is the overall scale which characterizes the threshold that noncommutative effects become relevant~\cite{das1,das2,das3}. The $c^{\mu \nu}$ matrix is analogous to the (electromagnetic) field tensor in structure. However, it is not at all a tensor because its elements are assumed to be constant in all reference frames. Before proceeding to numerical analysis let us make some general remarks regarding calculation of scattering amplitudes.

Firstly, two distinct cases should be discussed separately: The space-space or $B-$field type noncommutativity which means the elements $c^{ij}$ ($i,j$ run from 1 to 3) are non-vanishing and space-time or $E-$field type noncommutativity which means $c^{0j}$ elements are non-zero. Two types may have some features in common. The later type has been known to have some problems concerning the unitary and causality considerations~\cite{violation1,violation2}. Here, we will consider only the case of space-space noncommutativity.

Secondly, because that the vertex functions for minimal and non-minimal extended models are different, their phenomenological perspectives should be discussed separately. In evaluation of amplitudes we used~\eqref{eq:mWWZZ},~\eqref{eq:nmWWZZ} (based on the model under consideration) and also the relevant $\theta-$expanded rules developed in~\cite{melic1,betabi}. Notice that the amplitude of diagram~\ref{schannelhiggs} is equal in the minimal and non-minimal models because the Higgs couplings remain the same in both model extensions. This diagram may cause a distinction between two cases through interference contributions.

Lastly, as we mentioned earlier, noncommutative model introduces new interactions which are forbidden in the standard model. Those that are relevant to our discussion are $ZZZ$ and $\gamma ZZ$ couplings~\cite{melic1}. By using these vertices, it is possible to add diagrams~\ref{schannelweak},~\ref{schannelphoton} into standard Feynman graphs. Note that the later diagram is allowed only in the non-minimal noncommutative extension of the standard model. The scattering amplitudes of these diagrams are of order $(\theta^2)$ and their interference contributions are much important than their individual contributions to the total cross section.
\subsection{$B$-type noncommutativity}
Let us assume $\overrightarrow{\theta}_{\mathrm{B}} = \frac{1}{\sqrt{3}} (\hat{i} + \hat{j} + \hat{k})$ where $\theta^k_{\mathrm{B}} \equiv \epsilon^{ijk}c_{ij}$ for the case of $B$-field type noncommutativity. This is a constant vector and aligns in a specific direction in space (in all reference frames). As the earth rotates and revolves around the Sun, the direction of $\overrightarrow{\theta}_{\mathrm{B}}$ continuously changes and observable quantities are expected to show a specific time dependence. For those instants that our assumption on direction of $\overrightarrow{\theta}_{\mathrm{B}}$ is satisfied the $\frac{d\sigma_{\mathrm{NC}}}{d\phi}$ and $\frac{d\sigma_{\mathrm{NC}}}{d\cos{\theta}}$ distributions show a characteristic oscillatory behaviour. We have shown the numerical results of noncommutative standard model in Figs.~\ref{fig:figure4} -~\ref{fig:figure15} in Appendix~\ref{figure}. The figures in rows, from left to right, correspond respectively to $\frac{d\sigma_{\mathrm{NC}}}{d\phi}$ at $\theta=\frac{\pi}{2}$, the $\theta-$integrated $\frac{d\sigma_{\mathrm{NC}}}{d\phi}$ and $\phi-$integrated $\frac{d\sigma_{\mathrm{NC}}}{d\cos{\theta}}$ distributions for different values of $\Lambda_{\mathrm{NC}}$ and those in columns display the same quantity at different incident energies. Figure~\ref{fig:figure4}, exhibits the azimuthal distributions $\frac{d\sigma_{\mathrm{NC}}}{d\phi}$ at $\theta = \frac{\pi}{2}$ and $\sqrt{s} = 1.0$ TeV for $\Lambda_{\mathrm{NC}}=$ 0.6, 0.8, 1.2, 1.5, 1.8 TeV and $\infty$ in the context of minimal noncommutative model. Figure~\ref{fig:figure5}, represents the integrated $\frac{d\sigma_{\mathrm{NC}}}{d\phi}$ distributions and Fig.~\ref{fig:figure6} is the integrated $\frac{d\sigma_{\mathrm{NC}}}{d \cos{\theta}}$ distributions at the same c.m. energy for different scales. The evolution of differential cross sections by variation of noncommutativity scale can be easily understood from these graphics. We see that $\frac{d\sigma_{\mathrm{NC}}}{d\phi}$ distributions show an oscillatory behaviour with crests at $\phi=\frac{3\pi}{4},\frac{7\pi}{4}$. As the noncommutativity scale increases, the crests smoothly collapse and disappear at large enough scales. In particular, in the limit of $\Lambda_{\mathrm{NC}}=\infty$, we recover the results of the standard model (see blue curves in Figs.\ref{fig:figure1} -~\ref{fig:figure3}). The $\frac{d\sigma_{\mathrm{NC}}}{d \cos{\theta}}$ distributions show a symmetric pattern around $\theta=\frac{\pi}{2}$ with two local maxima at $\theta=\frac{\pi}{4},\frac{3\pi}{4}$ as in Fig.~\ref{fig:figure6}. The dashed curve is for $\Lambda_{\mathrm{NC}}=0.6$ TeV. Others are, however, hidden because they are tiny at $\sqrt{s}=1.0$ TeV. From phenomenological point of view the appearance of local maxima at $\theta=\frac{\pi}{4},\frac{3\pi}{4}$ means that particles are scattered much likely either in forward direction from $\theta=0$ to $\theta=\frac{\pi}{4}$ or in backward direction from $\theta=\frac{3\pi}{4}$ to $\theta=\pi$. Note that $\frac{d\sigma_{\mathrm{NC}}}{d \cos{\theta}}$ distributions are forward-backward symmetric as in the standard model. In Figs.~\ref{fig:figure7} -~\ref{fig:figure9}, we have shown the same distributions at $\sqrt{s} = 1.5$ TeV. The general features of azimuthal distributions are as those in Figs.~\ref{fig:figure4},~\ref{fig:figure5} except that in the present case crests are much sharp for $\Lambda_{\mathrm{NC}}=0.6$ while others are hidden. Observe that the local maxima for $\Lambda_{\mathrm{NC}}=$ 0.8 are visible in Fig.~\ref{fig:figure7}. By comparing the results we can conclude immediately that for a given scale $\Lambda_{\mathrm{NC}}$ as the center of mass energy increases crests become sharp and much strong (compare Figs.~\ref{fig:figure4} -~\ref{fig:figure6} respectively with Figs.~\ref{fig:figure5} -~\ref{fig:figure7}). Again, by increasing $\Lambda_{\mathrm{NC}}$, the oscillation amplitudes collapse and disappear as before.

Next, we consider the non-minimal noncommutative standard model. Figures~\ref{fig:figure10} -~\ref{fig:figure12}, display the numerical results of non-minimal model at $\sqrt{s} = 1.5$ TeV. The phenomenological implications of minimal and non-minimal models can be easily understood and compared using Figs.~\ref{fig:figure7} -~\ref{fig:figure12}. In the context of non-minimal model, azimuthal distributions upto a numerical factor of order $10^{5}$ are essentially the same as those in the minimal model. The $\frac{d\sigma_{\mathrm{NC}}}{d \cos{\theta}}$ distributions are, however, much different from similar distributions in the minimal case both in shape and scale. In this case, $\frac{d\sigma_{\mathrm{NC}}}{d \cos{\theta}}$ distributions show a strong peak at $\theta=\frac{\pi}{2}$ and outgoing $Z^0$ bosons are expected to be scattered most likely around $\theta=\frac{\pi}{2}$. Let us recall that $Z^0$ bosons are distributed symmetrically in forward and backward directions. Figurers~\ref{fig:figure13} -~\ref{fig:figure15}, exhibit the expected results of non-minimal model at $\sqrt{s} = 2.0$ TeV. Again, as $\Lambda_{\mathrm{NC}}$ increases, the characteristic oscillations of distributions are suppressed and disappear at $\Lambda_{\mathrm{NC}}=\infty$.
\subsection{Estimation of number of events in noncommutative model}
The number of events, i.e., the number of $W^+W^- \rightarrow Z Z$ (subprocess) scatterings, can be used to give a direct sense of implications of the noncommutative model. Assuming the integrated luminosity 100 $\mathrm{fb}^{-1}$, we estimated the number of signals in the context of standard model (SM), minimal noncommutative standard model (mNCSM) as well as the non-minimal model (nmNCSM) for some values of $\Lambda_{\mathrm{NC}}$ at c.m. energies $\sqrt{s}=$ 1.0, 1.5, and 2.0 TeV in Table~\ref{tab:tabel2}. The last three lines ($\Lambda_{\mathrm{NC}} = \infty$) correspond to predictions of the standard model.
\begin{table}[h!]
\centering
\begin{tabular}{llll}
\hline
\hline
Model & $\sqrt{s}$ (TeV) & \quad $\Lambda_{\mathrm{NC}}$ (TeV)  & \quad No. of events \\
\hline
      &      & \quad \quad 0.6 & \qquad 1.130 $\times$ 10$^7$ \\
mNCSM &  \quad 1.0 & \quad\quad 1.2 & \qquad 3.651 $\times$ 10$^6$ \\
      &      & \quad\quad 1.8 & \qquad 3.158 $\times$ 10$^6$\\
\hline
      &      & \quad\quad 0.6 & \qquad 1.921 $\times$ 10$^9$ \\
mNCSM &  \quad 1.5 & \quad\quad 1.2 & \qquad 1.080 $\times$ 10$^8$ \\
      &      & \quad\quad 1.8 & \qquad 6.200 $\times$ 10$^8$ \\
\hline
       &     & \quad\quad 0.6 & \qquad 5.543 $\times$ 10$^{12}$ \\
nmNCSM & \quad 1.5 & \quad\quad 1.2 & \qquad 1.790 $\times$ 10$^{11}$ \\
       &     & \quad\quad 1.8 & \qquad 1.036 $\times$ 10$^{11}$ \\
\hline
       &     & \quad\quad 0.6 & \qquad 1.054 $\times$ 10$^{13}$ \\
nmNCSM & \quad 2.0 & \quad\quad 1.2 & \qquad 3.427 $\times$ 10$^{12}$ \\
       &     & \quad\quad 1.8 & \qquad 1.972 $\times$ 10$^{11}$ \\
\hline
       & \quad 1.0   & \quad\quad $\infty$ & \qquad 2.960 $\times$ 10$^6$ \\
SM     & \quad 1.5   & \quad\quad $\infty$ & \qquad 2.862 $\times$ 10$^6$ \\
       & \quad 2.0   & \quad\quad $\infty$ & \qquad 2.763 $\times$ 10$^6$ \\
\hline
\hline
\end{tabular}
\caption{Number of signals in mNCSM, nmNCSM and SM at integrated luminosity 100 fb$^{-1}$.}
\label{tab:tabel2}
\end{table}
\section{Summary and Conclusion}\label{conclu}
We examined the gauge sector of both the minimal and non-minimal noncommutative standard model and obtained the $\mathcal{O}\, (\theta)$ Feynman rules for all QGC's. It was found that the Higgs part of the action induces contributions into electroweak gauge sector of noncommutative standard model. In the minimal case and upto the leading order of deformation quantity, the electroweak gauge sector of the model is the same as that of the standard model and only the Higgs sector induced interactions contribute to $\mathcal{O}\, (\theta)$ Feynman rules for gauge boson couplings. These contributions are of dimension-4 and momentum independent. In contrast, in the non-minimal case the gauge sector of the model contributes to QGC's through dimension-6 and momentum dependent interactions. Also, two anomalous couplings appear in the non-minimal model where photon as well as the neutral weak boson are coupled directly to three gluons. Such an electroweak-chromodynamics mixing is forbidden in the standard model at tree level. We studied the phenomenological implications of the model in $W^-W^+\rightarrow ZZ$ scattering and showed that noncommutativity od spacetime manifest itself through a characteristic oscillatory behaviour in azimuthal distribution of differential cross sections. In particular, for the case of space-space noncommutativity we evaluated scattering cross sections at $\sqrt{s}=1.0, \, 1.5, \, 2.0$ TeV for $\Lambda_{\mathrm{NC}}$ from 0.6 to 1.8 TeV and found that the $\frac{d\sigma_{\mathrm{NC}}}{d\phi}$ distributions at $\theta=\frac{\pi}{2}$ as well as the integrated $\frac{d\sigma_{\mathrm{NC}}}{d\phi}$ distributions show a sinusoidal behaviour with crests at $\phi = \frac{3\pi}{4}, \frac{7\pi}{4}$. For a given c.m. energy as $\Lambda_{\mathrm{NC}}$ increases crests smoothly collapse and disappear at large scales. On the other hand, for a fixed value of $\Lambda_{\mathrm{NC}}$ by increasing the c.m. energy crests become much strong and appear as sharp peaks. Also, the $\frac{d\sigma_{\mathrm{NC}}}{d \cos{\theta}}$ distributions show a symmetric pattern around $\theta = \frac{\pi}{2}$. However, the patterns for minimal and non-minimal models are different in shape. In the minimal model, two separate crests appear at $\theta = \frac{\pi}{4}, \frac{3\pi}{4}$ while in the non-minimal case there is a central peak at $\theta = \frac{\pi}{2}$ and curves smoothly disappear at forward-backward directions. Analysis of $\frac{d\sigma_{\mathrm{NC}}}{d \cos{\theta}}$ distributions indicate that in minimal model the number of events, in comparison with the standard model, increases considerably in backward direction from $\theta = 0$ to $\theta=\frac{\pi}{4}$ and also in forward direction from $\theta=\frac{3\pi}{4}$ to $\theta=\pi$ while from $\theta=\frac{\pi}{4}$ to $\theta=\frac{3\pi}{4}$ remains essentially the same. In contrast, in the non-minimal case the number of events is expected to increase from $\theta=\frac{\pi}{4}$ to $\theta=\frac{3\pi}{4}$ with a sharp maximum at $\theta=\frac{\pi}{2}$. However, in both the models the scattering is forward-backward symmetric. Assuming the integrated luminosity 100 $\mathrm{fb}^{-1}$, we estimated the number of $W^-W^+\rightarrow ZZ$ scatterings in both the minimal and non-minimal noncommutative models for some values of $\Lambda_{\mathrm{NC}}$ at c.m. energies $\sqrt{s}=1.0, \, 1.5, \, 2.0$ TeV and compared the results with predictions of the standard model. The number of events are expected to increase by a factor of order 10$^1$ upto 10$^5$ in some cases.
\appendix
\section{Feynman diagrams}\label{diag}
In the standard model, there are four diagrams which contribute to $W^-W^+\rightarrow ZZ$ scattering. These are contact coupling, $t$-channel, $u$-channel and the Higgs mediated $s$-channel diagrams. In the context of noncommutative standard model the scattering amplitude of these diagrams are evaluated using the $\theta$-expanded vertex functions. Also, two new diagrams~\ref{schannelweak} and~\ref{schannelphoton} are allowed in noncommutative extended model. Oserve that, the last diagram contributes only in non-minimal model.
%
%
\begin{enumerate}
\begin{minipage}[H]{2.2 in}
\item contact\label{contact}
  \\
  \\
 \tikzset{
particle/.style={decorate, draw=black, decoration={snake=coil}},
gluon/.style={decorate, draw=black, decoration={coil,aspect=1.2}},
}
\begin{tikzpicture}
[node distance=1.0cm and 1.0cm]
\coordinate[label=left:$W_\nu^{+}$] (left up);
\coordinate[below right=of left up] (mid);
\coordinate[above right=of mid,label=right:$Z_\lambda$] (right up);
\coordinate[below left=of mid,label=left:$W_\mu^{-}$] (left down);
\coordinate[below right=of mid,label=right:$Z_\kappa$] (right down);
\draw[particle] (left up) -- (mid);
\draw[particle] (right up) -- (mid);
\draw[particle] (left down) -- (mid);
\draw[particle] (right down) -- (mid);
\end{tikzpicture}
\end{minipage}
\begin{minipage}[H]{2.2 in}
\item $t\,$- channel\label{tchannel}
 \\
 \\
 \tikzset{
particle/.style={decorate, draw=black, decoration={snake=coil}},
gluon/.style={decorate, draw=black, decoration={coil,aspect=1.2}},
}
\begin{tikzpicture}
[node distance=1.2cm and 1.2cm]
\coordinate[label=left:$W_\nu^{+}$] (left up);
\coordinate[right=of left up] (midup);
\coordinate[below=2.0cm of midup] (middown);
\coordinate[right=of midup,label=right:$Z_\lambda$] (right up);
\coordinate[left=of middown,label=left:$W_\mu^{-}$] (left down);
\coordinate[right=of middown,label=right:$Z_\kappa$] (right down);
\draw[particle] (left up) -- (midup);
\draw[particle] (right up) -- (midup);
\draw[particle] (midup) -- (middown);
\draw[particle] (left down) -- (middown);
\draw[particle] (right down) -- (middown);
\end{tikzpicture}
\end{minipage}
\begin{minipage}[H]{2.2 in}
\item $u\,$- channel,\label{uchannel}
 \\
 \\
 \tikzset{
particle/.style={decorate, draw=black, decoration={snake=coil}},
gluon/.style={decorate, draw=black, decoration={coil,aspect=1.2}},
}
\begin{tikzpicture}
[node distance=1.2cm and 1.2cm]
\coordinate[label=left:$W_\nu^{+}$] (left up);
\coordinate[right=of left up] (midup);
\coordinate[below=2.0cm of midup] (middown);
\coordinate[right=of midup,label=right:$Z_\kappa$] (right up);
\coordinate[left=of middown,label=left:$W_\mu^{-}$] (left down);
\coordinate[right=of middown,label=right:$Z_\lambda$] (right down);
\draw[particle] (left up) -- (midup);
\draw[particle] (right down) -- (midup);
\draw[particle] (midup) -- (middown);
\draw[particle] (left down) -- (middown);
\draw[particle] (right up) -- (middown);
\end{tikzpicture}
\end{minipage}
\\
\\
\\
\begin{minipage}[H]{2.2 in}
\item $s\,$- channel,\label{schannelhiggs}
 \\
 \\
 \tikzset{
particle/.style={decorate, draw=black, decoration={snake=coil}},
gluon/.style={decorate, draw=black, decoration={coil,aspect=1.2}},
}
\begin{tikzpicture}
[node distance=1.2cm and 1.2cm]
\coordinate[label=left:$W_\nu^{+}$] (left up);
\coordinate[below=of left up] (midleft);
\coordinate[right=0.9cm of midleft,label=above:$\mathrm{H}$] (mid);
\coordinate[right=1.8cm of midleft] (midright);
\coordinate[above=of midright,label=right:$Z_\lambda$] (right up);
\coordinate[below=of midleft,label=left:$W_\mu^{-}$] (left down);
\coordinate[below=of midright,label=right:$Z_\kappa$] (right down);
\draw[particle] (left up) -- (midleft);
\draw[particle] (left down) -- (midleft);
\draw[dashed] (midleft) -- (midright);
\draw[particle] (right up) -- (midright);
\draw[particle] (right down) -- (midright);
\end{tikzpicture}
\end{minipage}
\begin{minipage}[H]{2.2 in}
\item $s\,$- channel,\label{schannelweak}
 \\
 \\
 \tikzset{
particle/.style={decorate, draw=black, decoration={snake=coil}},
gluon/.style={decorate, draw=black, decoration={coil,aspect=1.2}},
}
\begin{tikzpicture}
[node distance=1.2cm and 1.2cm]
\coordinate[label=left:$W_\nu^{+}$] (left up);
\coordinate[below=of left up] (midleft);
\coordinate[right=0.9cm of midleft,label=above:$Z$] (mid);
\coordinate[right=1.8cm of midleft] (midright);
\coordinate[above=of midright,label=right:$Z_\lambda$] (right up);
\coordinate[below=of midleft,label=left:$W_\mu^{-}$] (left down);
\coordinate[below=of midright,label=right:$Z_\kappa$] (right down);
\draw[particle] (left up) -- (midleft);
\draw[particle] (left down) -- (midleft);
\draw[particle] (midleft) -- (midright);
\draw[particle] (right up) -- (midright);
\draw[particle] (right down) -- (midright);
\end{tikzpicture}
\end{minipage}
\begin{minipage}[H]{2.2 in}
\item $s\,$- channel,\label{schannelphoton}
 \\
 \\
 \tikzset{
particle/.style={decorate, draw=black, decoration={snake=coil}},
gluon/.style={decorate, draw=black, decoration={coil,aspect=1.2}},
}
\begin{tikzpicture}
[node distance=1.2cm and 1.2cm]
\coordinate[label=left:$W_\nu^{+}$] (left up);
\coordinate[below=of left up] (midleft);
\coordinate[right=0.9cm of midleft,label=above:$\gamma$] (mid);
\coordinate[right=1.8cm of midleft] (midright);
\coordinate[above=of midright,label=right:$Z_\lambda$] (right up);
\coordinate[below=of midleft,label=left:$W_\mu^{-}$] (left down);
\coordinate[below=of midright,label=right:$Z_\kappa$] (right down);
\draw[particle] (left up) -- (midleft);
\draw[particle] (left down) -- (midleft);
\draw[particle] (midleft) -- (midright);
\draw[particle] (right up) -- (midright);
\draw[particle] (right down) -- (midright);
\end{tikzpicture}
\end{minipage}
\end{enumerate}
\section{Figures}\label{figure}
\begin{figure}[H]
\captionwidth 2.0in
\begin{minipage}[H]{2.2 in}
\includegraphics[width=2.0 in]{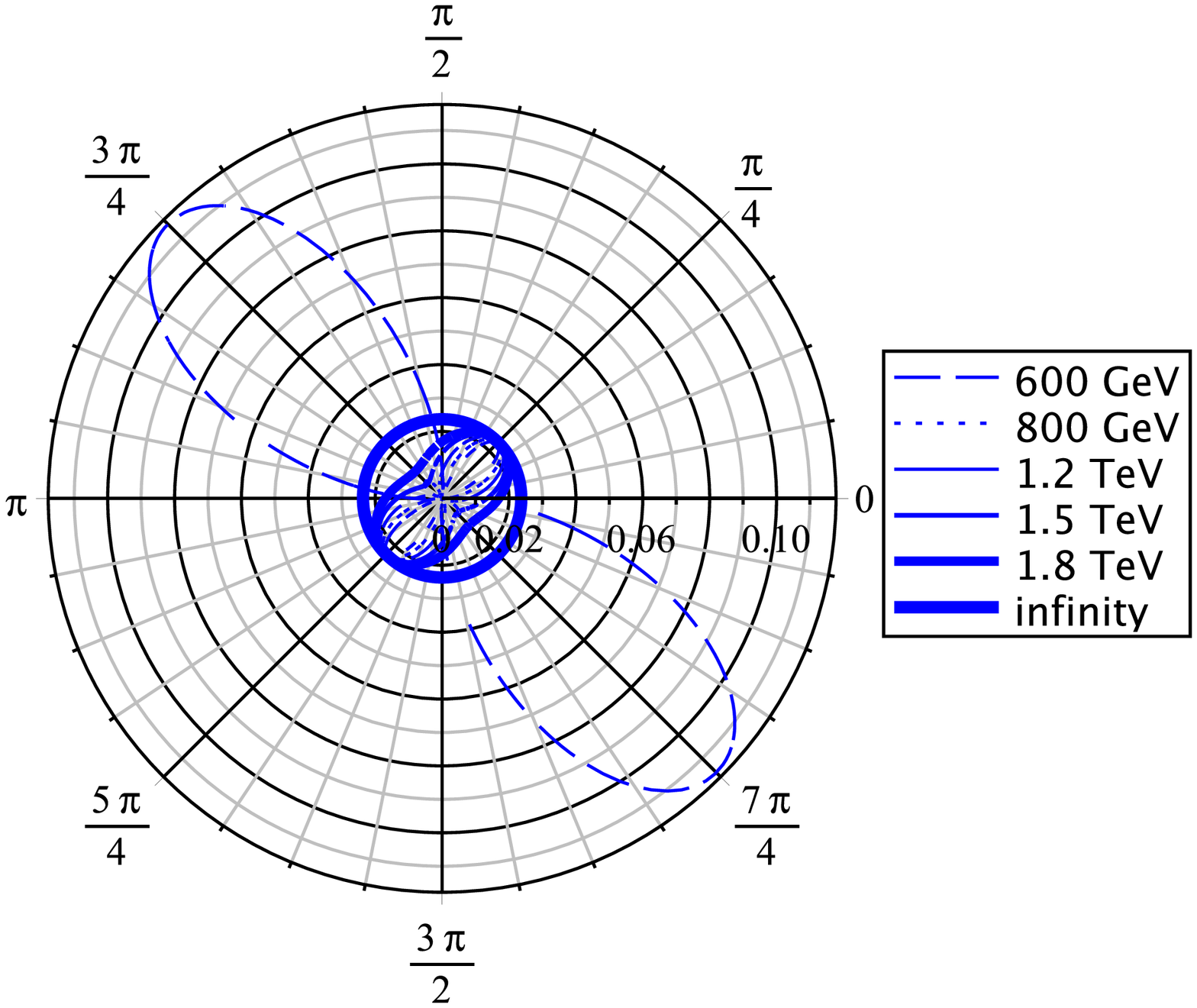}
\caption{Differential distributions $\frac{d\sigma_{\mathrm{NC}}}{d\phi}$ at ${\theta=\frac{\pi}{2}}$ in mNCSM  for $\sqrt{s}=1.0$ TeV.}
\label{fig:figure4}
\end{minipage}
\begin{minipage}[H]{2.2 in}
\includegraphics[width=2.0 in]{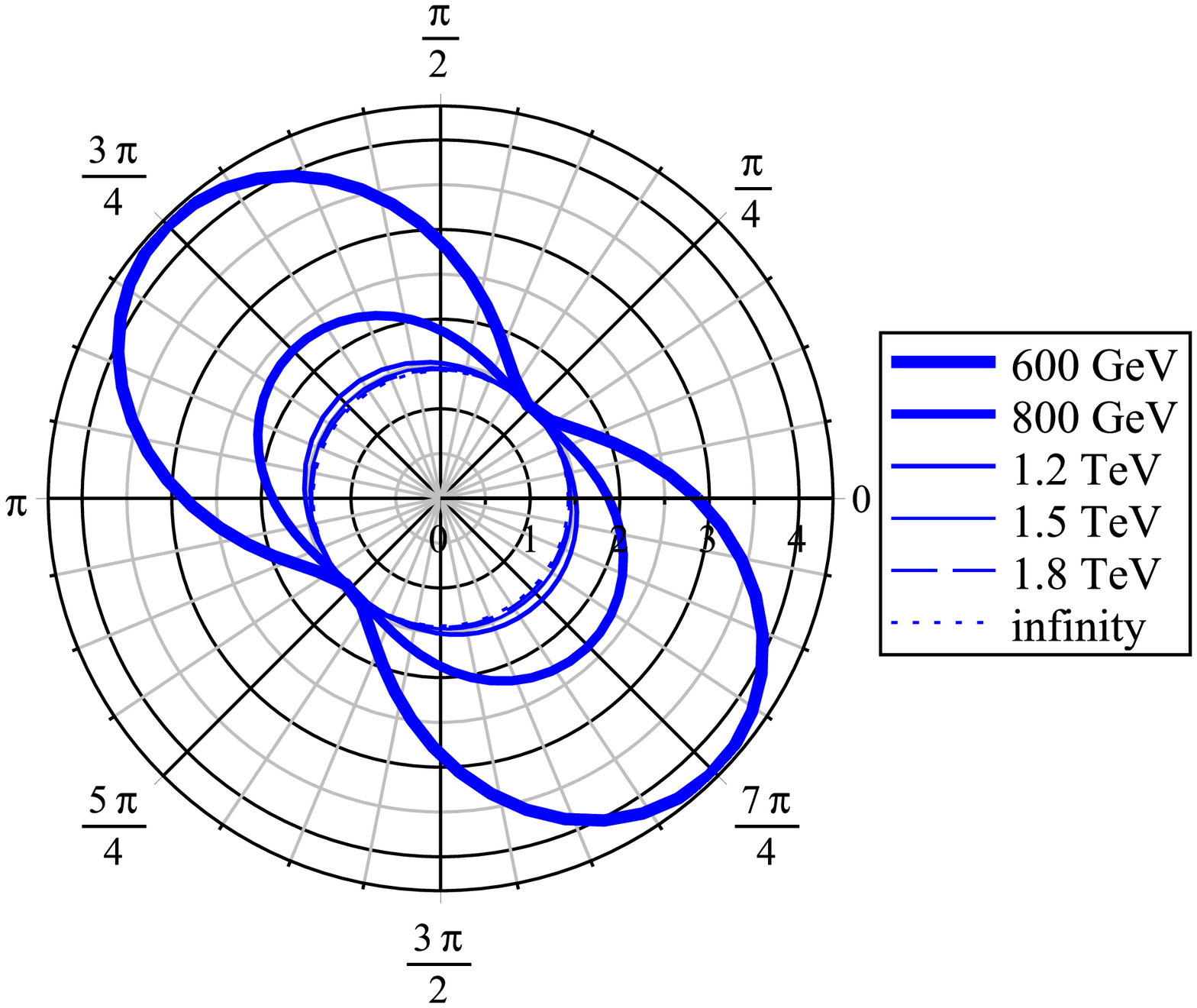}
\caption{The $\theta-$integrated $\frac{d\sigma_{\mathrm{NC}}}{d\phi}$ distributions in mNCSM  for $\sqrt{s}=1.0$ TeV.}
\label{fig:figure5}
\end{minipage}
\begin{minipage}[H]{2.2 in}
\includegraphics[width=2.0 in]{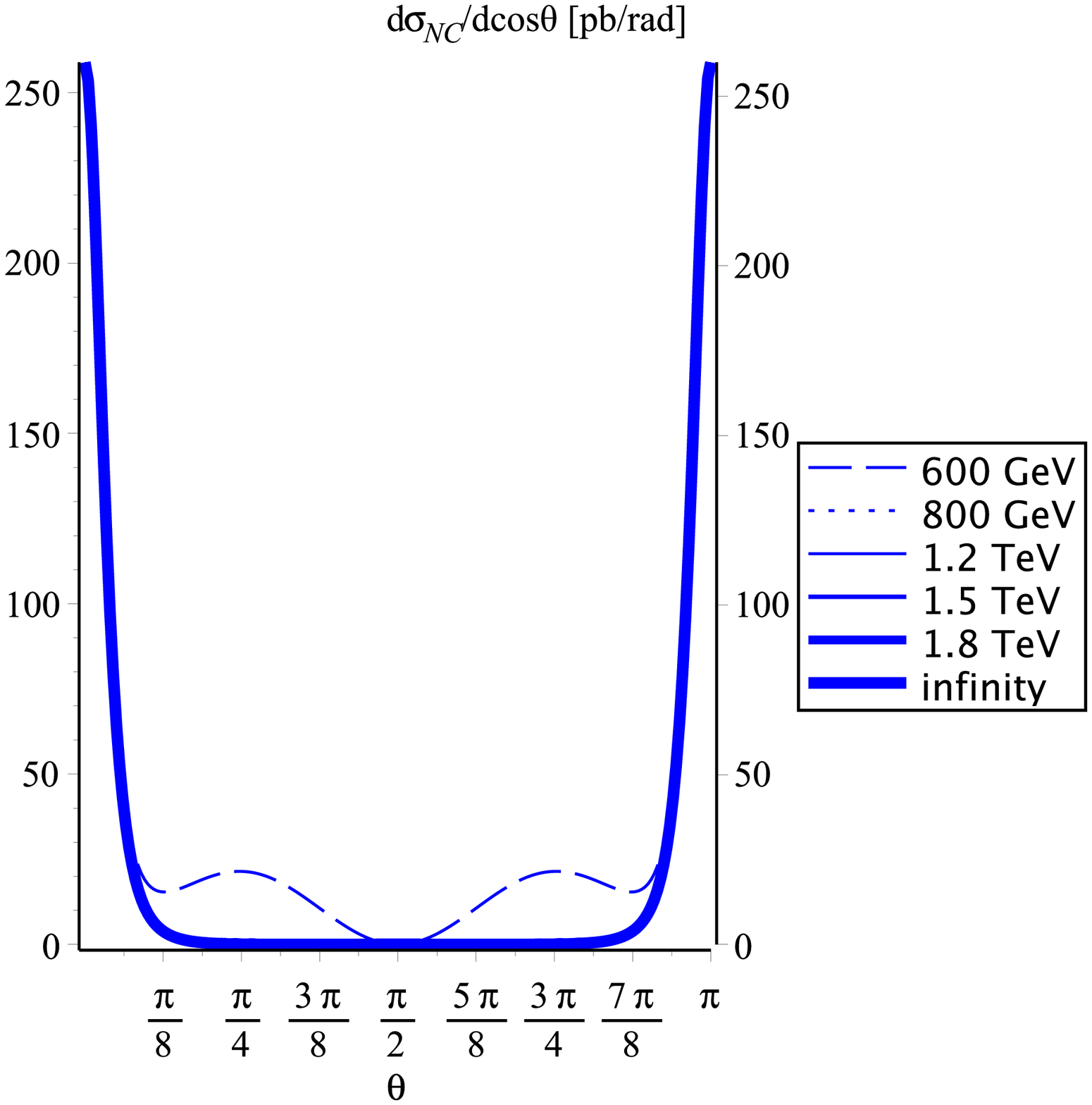}
\caption{The $\phi-$integrated $\frac{d\sigma_{\mathrm{NC}}}{d\cos{\theta}}$ distributions in mNCSM  for $\sqrt{s}=1.0$ TeV.}
\label{fig:figure6}
\end{minipage}
\end{figure}
\begin{figure}[H]
\captionwidth 2.0 in
\begin{minipage}[H]{2.2 in}
\includegraphics[width=2.0 in]{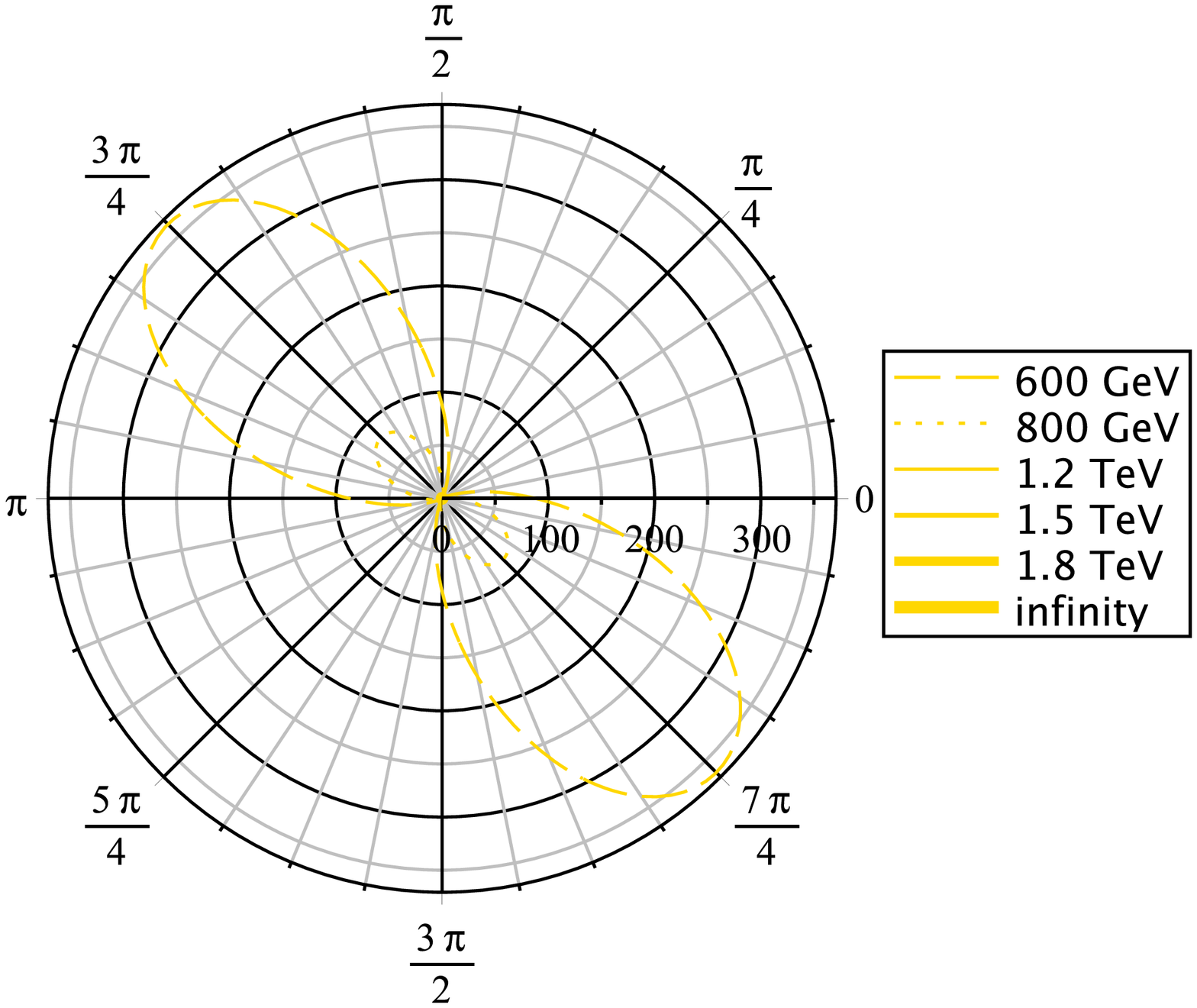}
\caption{Differential distributions $\frac{d\sigma_{\mathrm{NC}}}{d\phi}$ at ${\theta=\frac{\pi}{2}}$ in mNCSM for $\sqrt{s}=1.5$ TeV.}
\label{fig:figure7}
\end{minipage}
\begin{minipage}[H]{2.2 in}
\includegraphics[width=2.0 in]{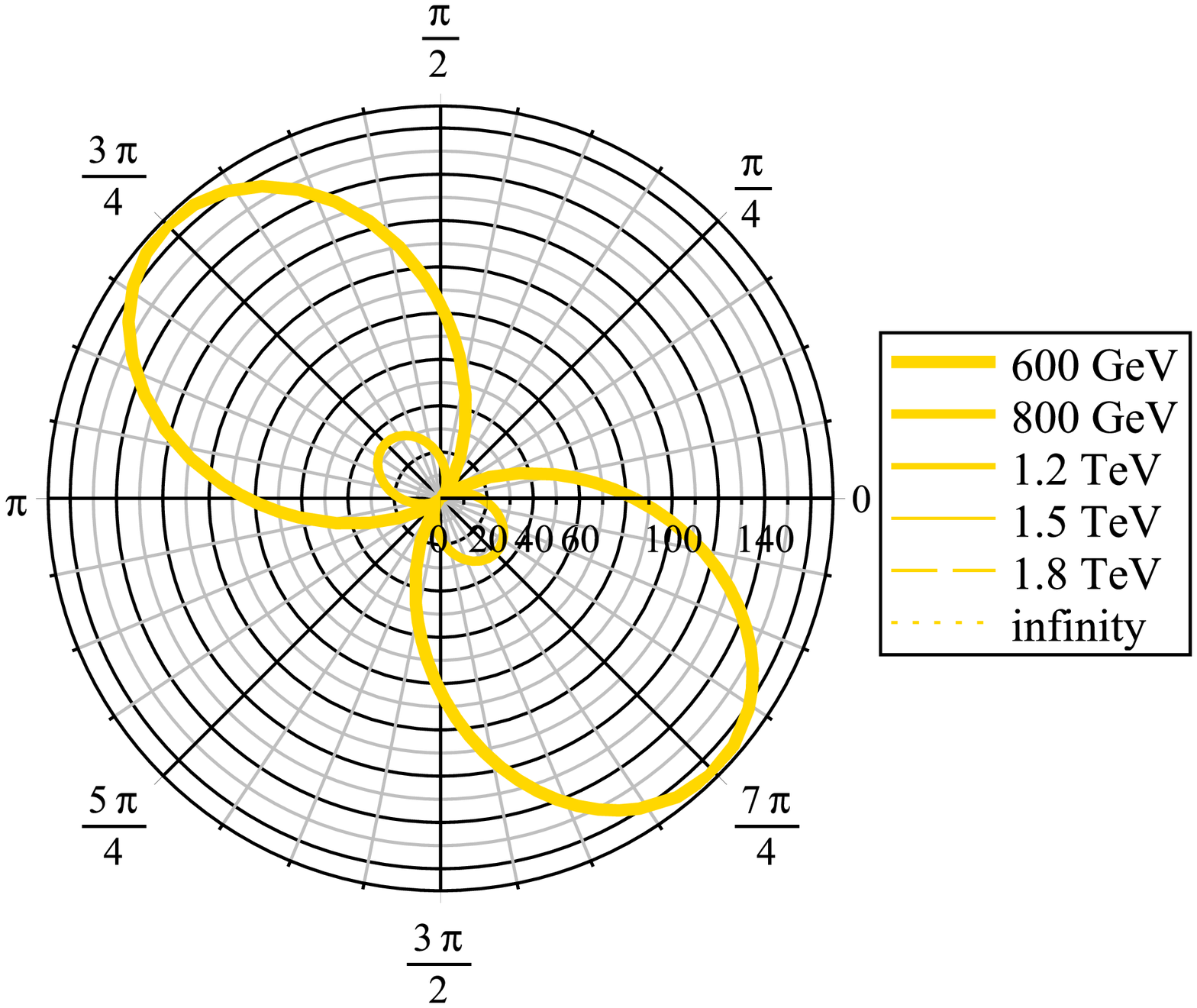}
\caption{The $\theta-$integrated $\frac{d\sigma_{\mathrm{NC}}}{d\phi}$ distributions in mNCSM for $\sqrt{s}=1.5$ TeV.}
\label{fig:figure8}
\end{minipage}
\begin{minipage}[H]{2.2 in}
\includegraphics[width=2.0 in]{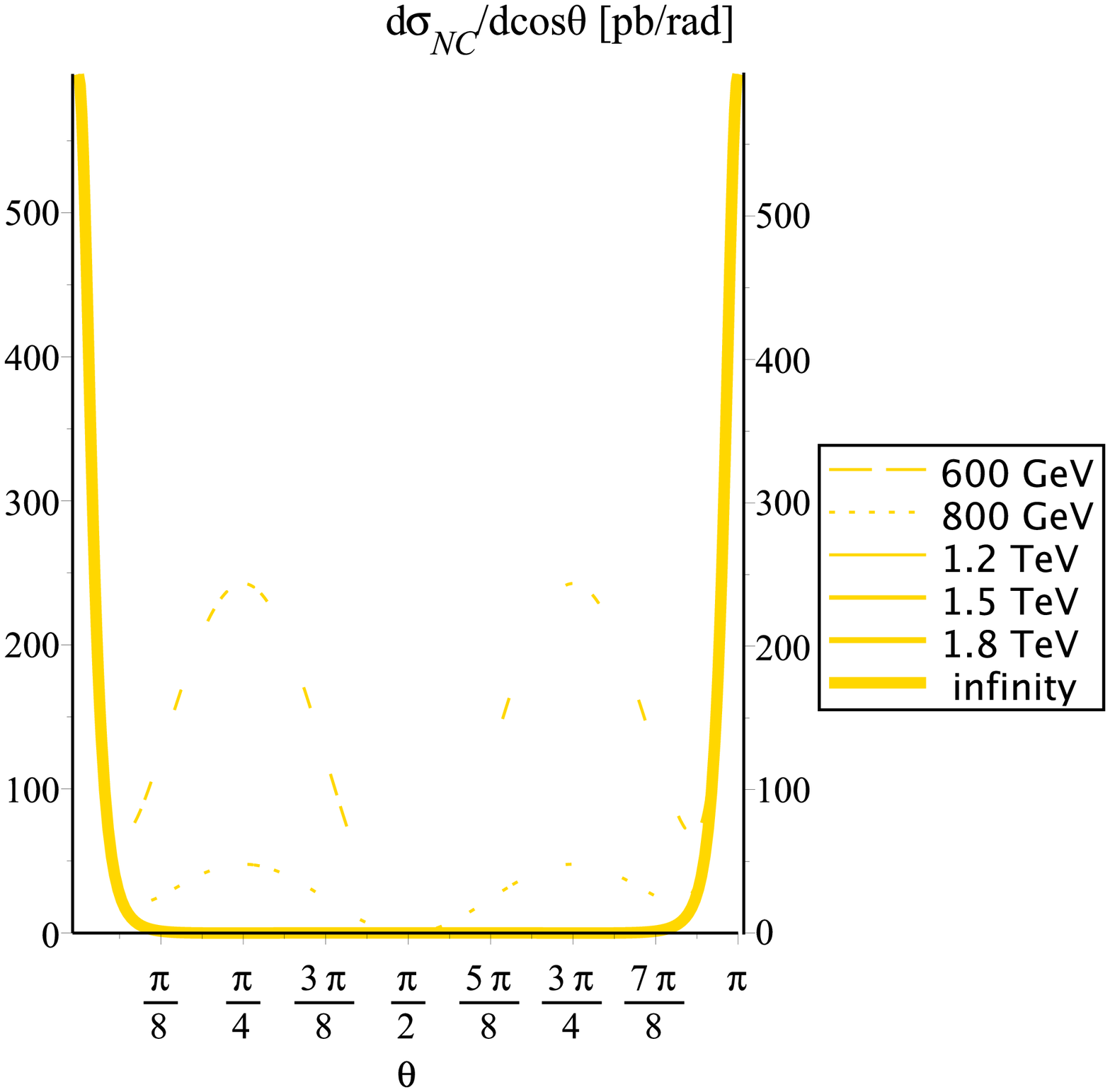}
\caption{The $\phi-$integrated $\frac{d\sigma_{\mathrm{NC}}}{d\cos{\theta}}$ distributions in mNCSM for $\sqrt{s}=1.5$ TeV.}
\label{fig:figure9}
\end{minipage}
\end{figure}
\begin{figure}[H]
\captionwidth 2.0 in
\begin{minipage}[H]{2.2 in}
\includegraphics[width=2.0 in]{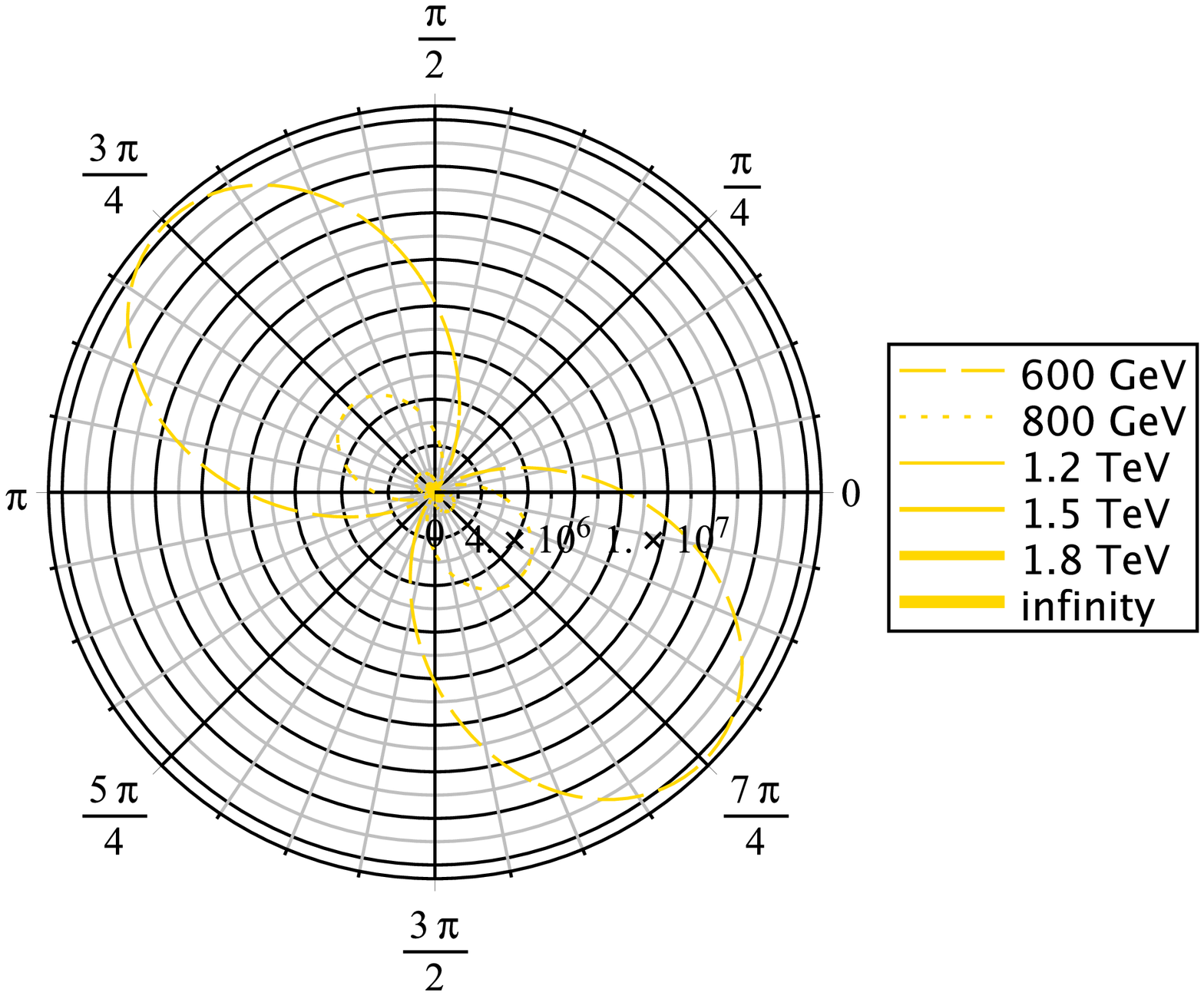}
\caption{Differential distributions $\frac{d\sigma_{\mathrm{NC}}}{d\phi}$ at ${\theta=\frac{\pi}{2}}$ in nmNCSM for $\sqrt{s}=1.5$ TeV.}
\label{fig:figure10}
\end{minipage}
\begin{minipage}[H]{2.2 in}
\includegraphics[width=2.0 in]{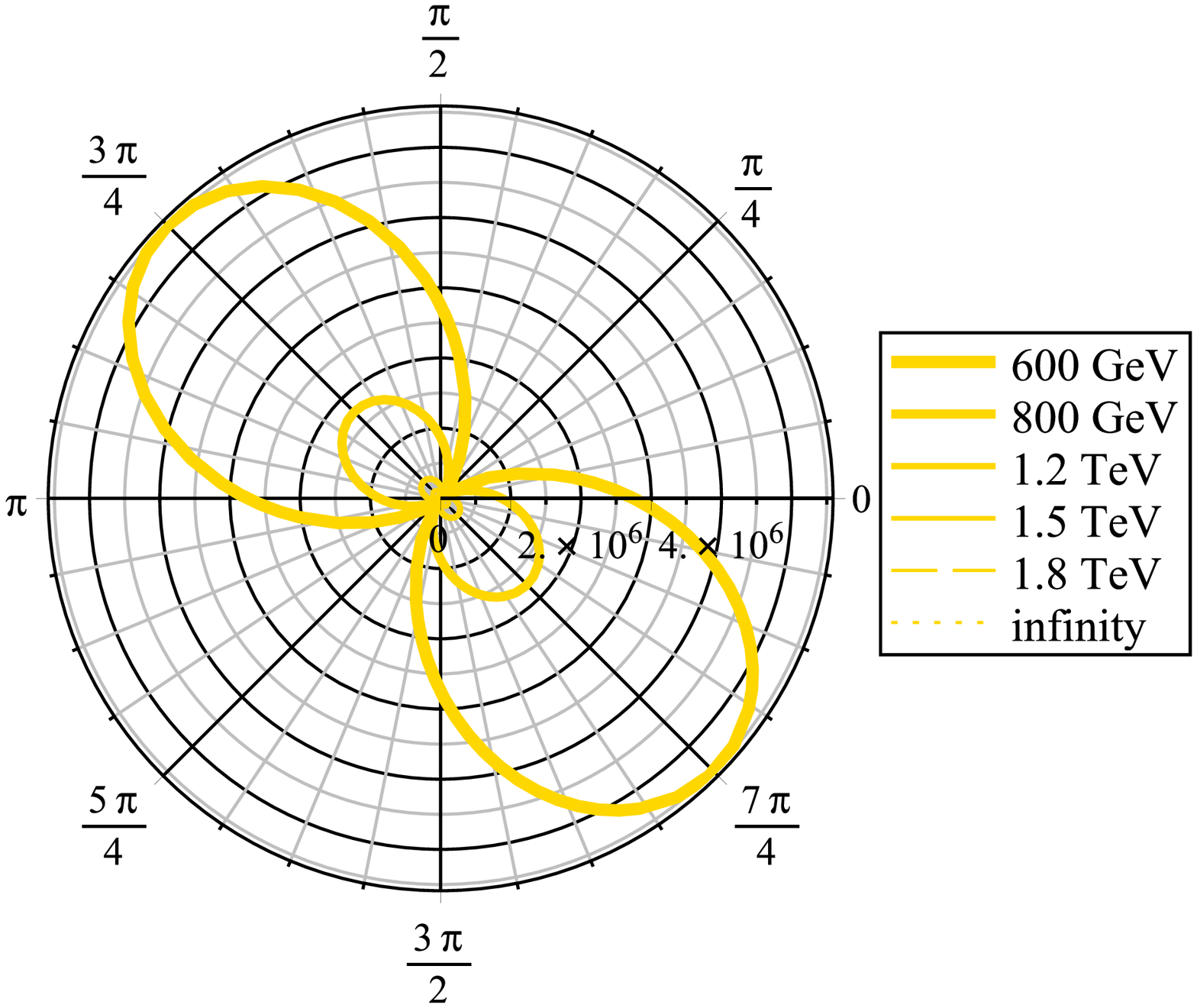}
\caption{The $\theta-$integrated $\frac{d\sigma_{\mathrm{NC}}}{d\phi}$ distributions in nmNCSM for $\sqrt{s}=1.5$ TeV.}
\label{fig:figure11}
\end{minipage}
\begin{minipage}[H]{2.2 in}
\includegraphics[width=2.0 in]{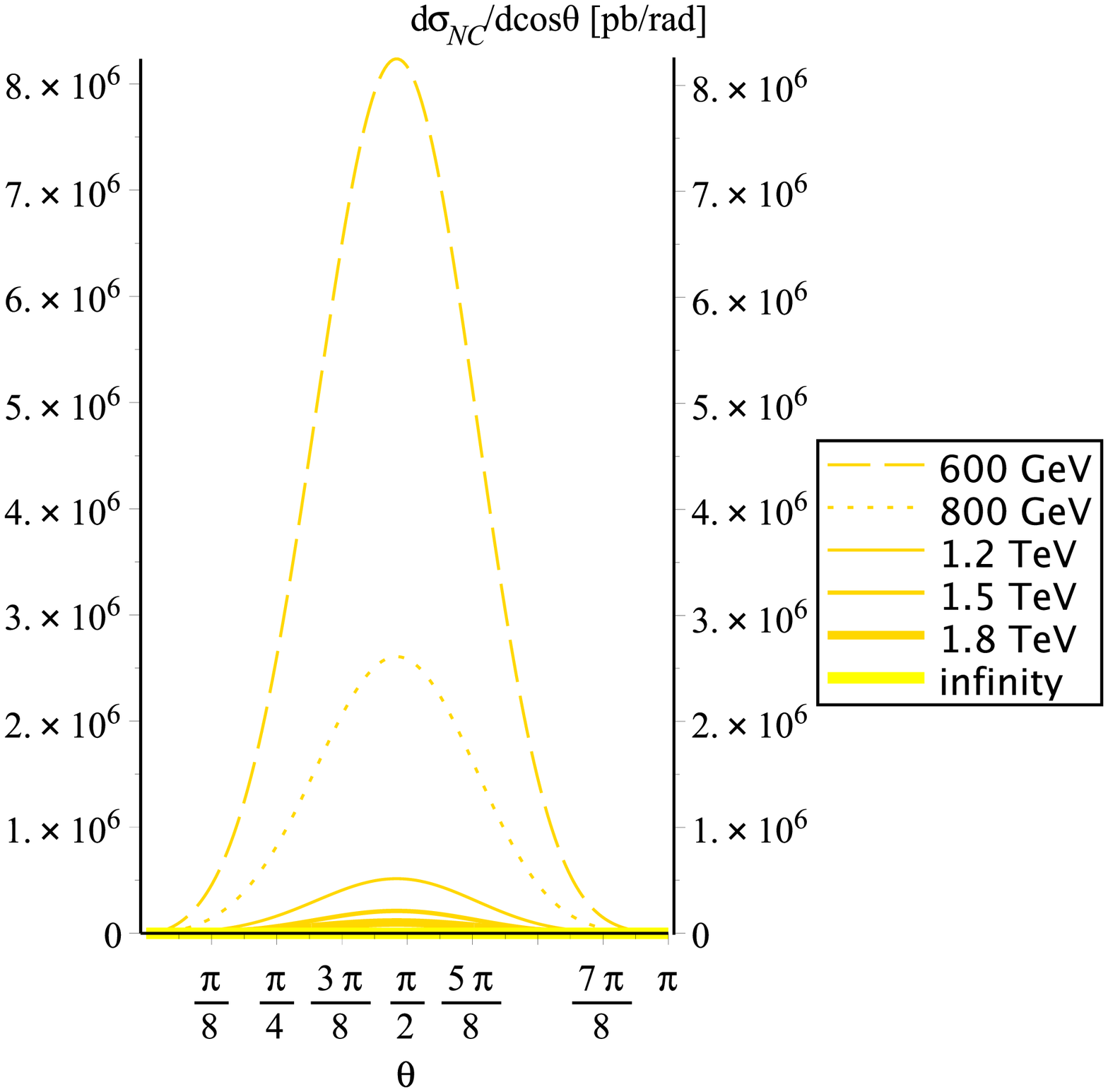}
\caption{The $\phi-$integrated $\frac{d\sigma_{\mathrm{NC}}}{d\cos{\theta}}$ distributions in nmNCSM for $\sqrt{s}=1.5$ TeV.}
\label{fig:figure12}
\end{minipage}
\end{figure}
\begin{figure}[H]
\captionwidth 2.0 in
\begin{minipage}[H]{2.2 in}
\includegraphics[width=2.0 in]{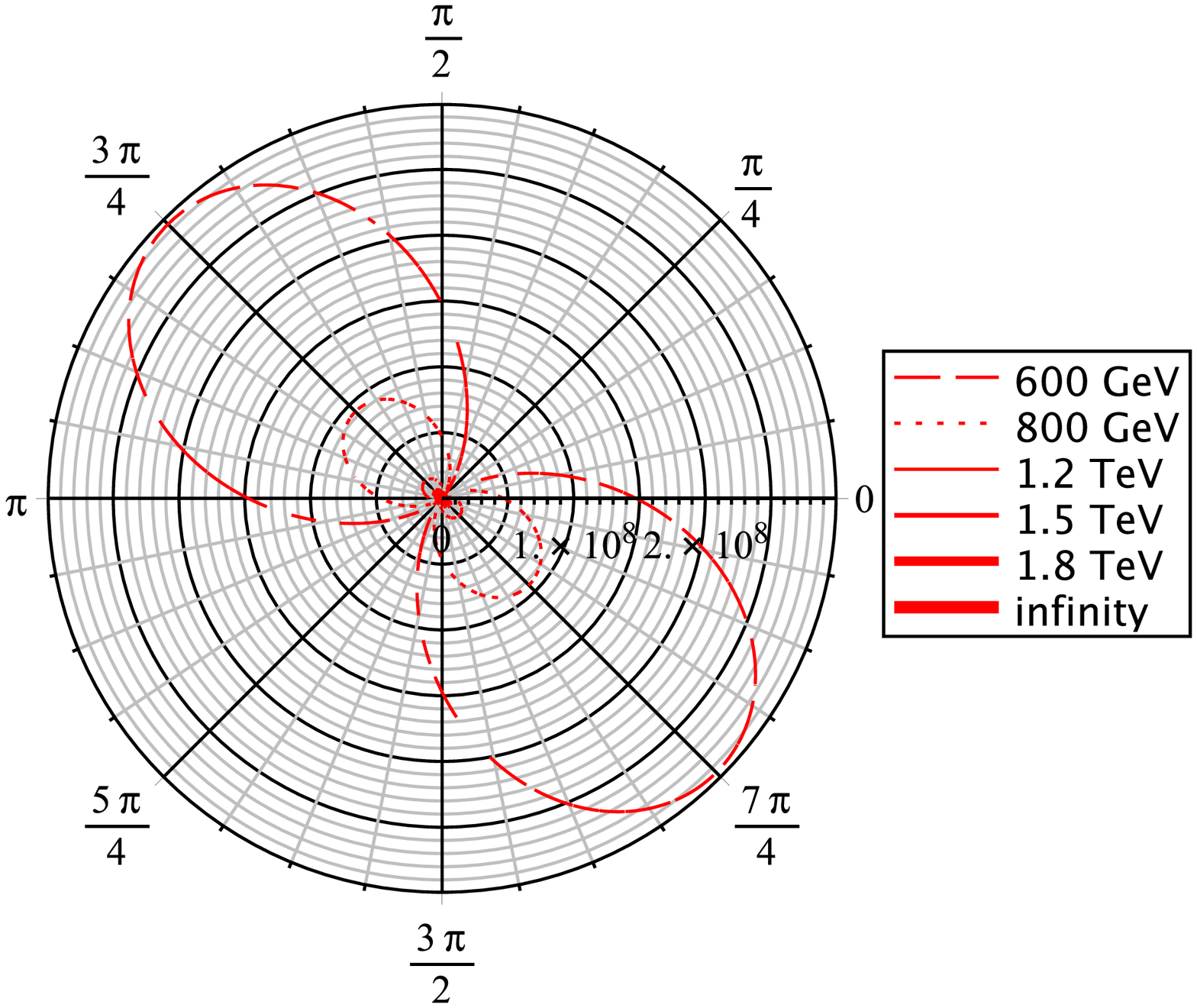}
\caption{Differential distributions $\frac{d\sigma_{\mathrm{NC}}}{d\phi}$ at ${\theta=\frac{\pi}{2}}$ in nmNCSM for $\sqrt{s}=2.0$ TeV.}
\label{fig:figure13}
\end{minipage}
\begin{minipage}[H]{2.2 in}
\includegraphics[width=2.0 in]{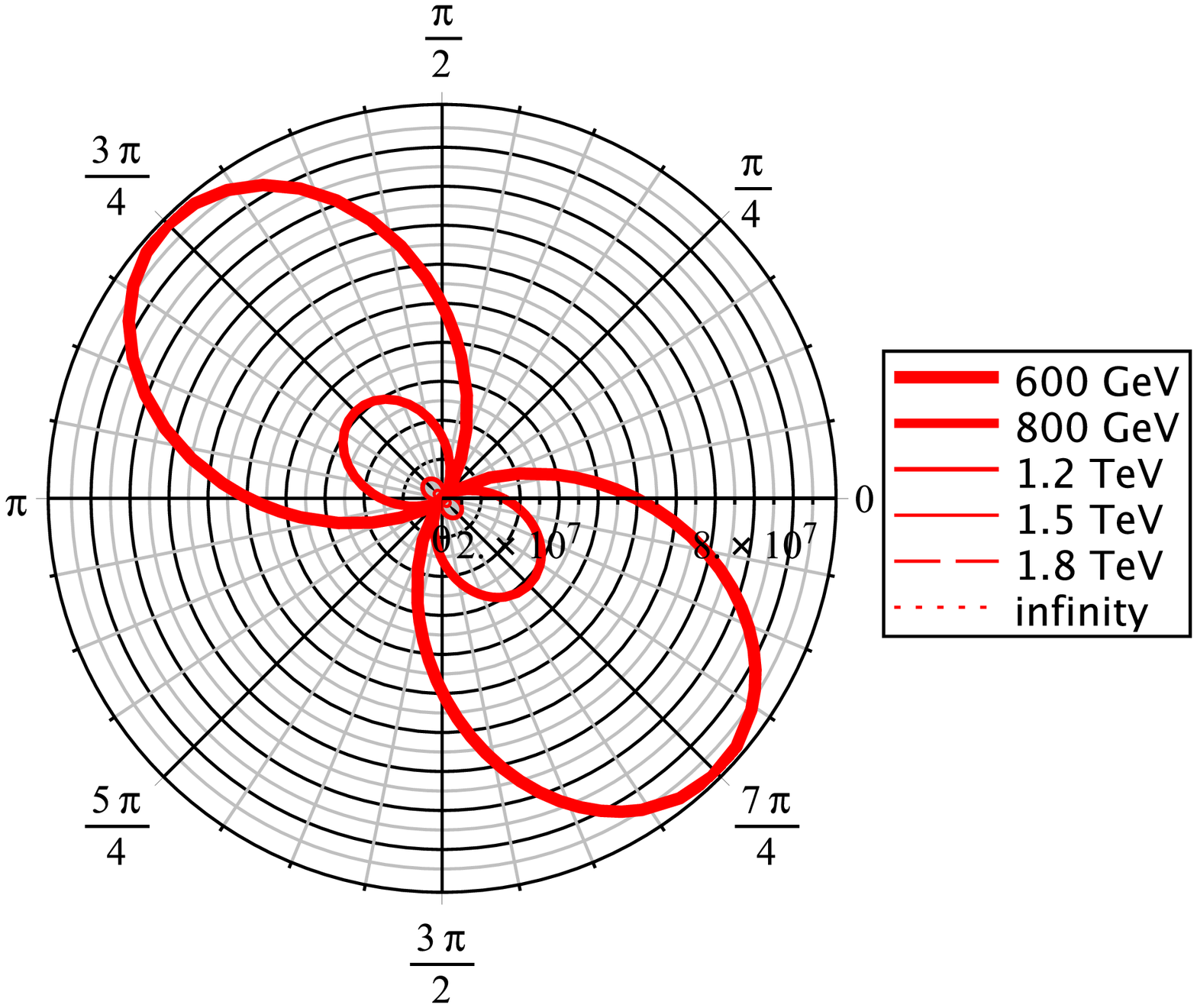}
\caption{The $\theta-$integrated $\frac{d\sigma_{\mathrm{NC}}}{d\phi}$ distributions in nmNCSM for $\sqrt{s}=2.0$ TeV.}
\label{fig:figure14}
\end{minipage}
\begin{minipage}[H]{2.2 in}
\includegraphics[width=2.0 in]{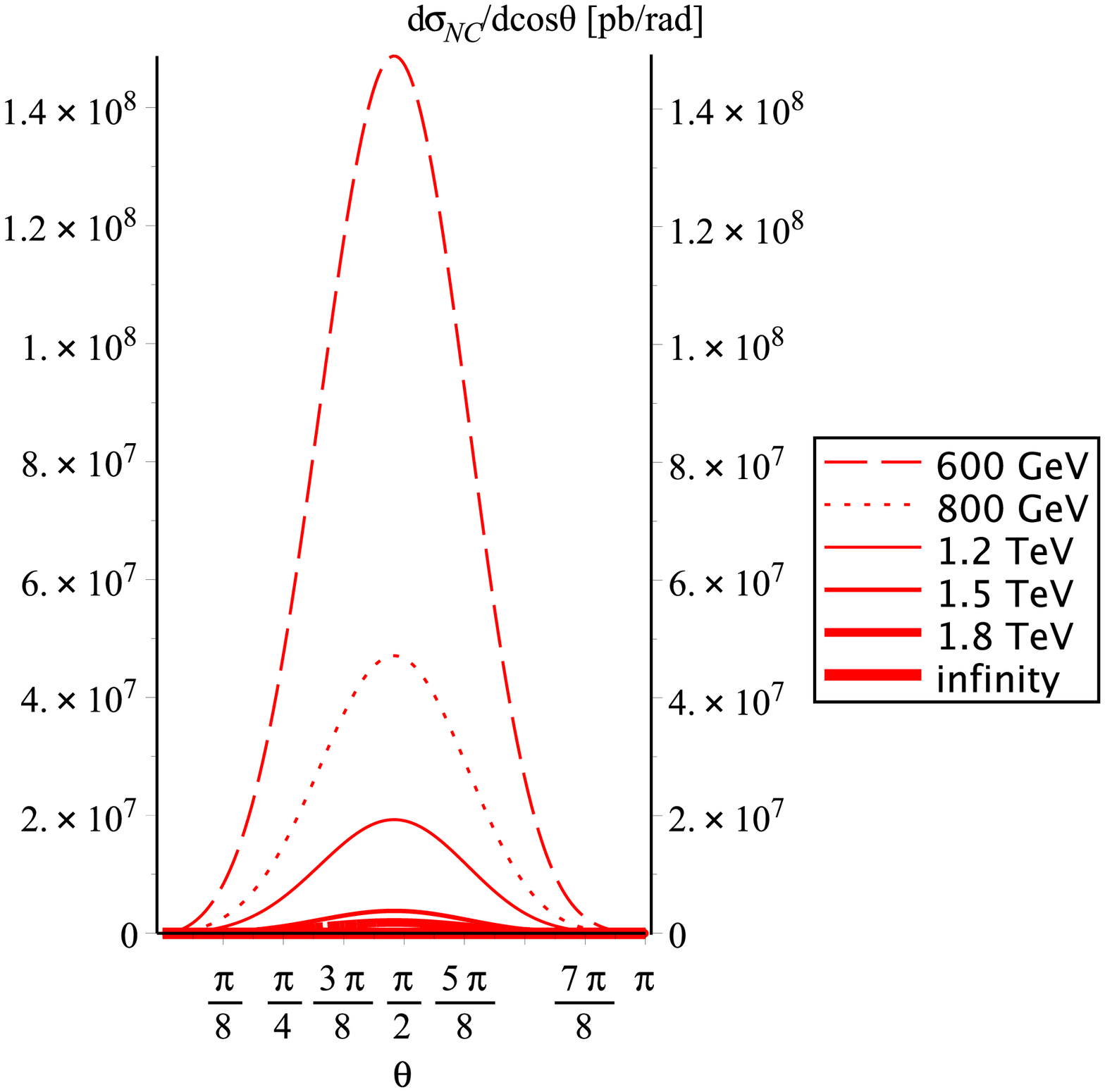}
\caption{The $\phi-$integrated $\frac{d\sigma_{\mathrm{NC}}}{d\cos{\theta}}$ distributions in nmNCSM for $\sqrt{s}=2.0$ TeV.}
\label{fig:figure15}
\end{minipage}
\end{figure}
%


\begin{thebibliography}{99}

\bibitem{sleva} J. Selvaganapathy, P.K. Das, and P. Konar, Phys. Rev. D {\bf93}, 116003 (2016).

\bibitem{ahmad} N. Ahmadiniaz, O. Corradini, D. \'{D}Ascanio, S. Estrada-Jim\'{e}nez, and P. Pisani, J. High. Energy. Phys. {\bf11}, 069 (2015).

\bibitem{chegal} M. Moumni and A. Benslama, Int. J. Mod. Phys. A {\bf28}, 1350139 (2013).
   				
\bibitem{mansour} S. Aghababaei, M. Haghighat, and A. Kheirandish, Phys. Rev. D {\bf87}, 047703 (2013).

\bibitem{ettefagh} M.M. Ettefaghi and M. Haghighat, Phys. Rev. D {\bf77}, 056009 (2008).

\bibitem{anaa} A. Alboteanu, T. Ohl and R. R\"{o}ckel, Phys. Rev. D {\bf74}, 096004 (2006).

\bibitem{ohl} T. Ohl and J. Reuter, Phys. Rev. D {\bf70}, 076007 (2004).

\bibitem{muller} L. M\"{o}ller, J. High. Energy. Phys. {\bf10}, 063 (2004).

\bibitem{seiberg} N. Seiberg and E. Witten, J. High. Energy. Phys. {\bf09}, 032 (1999).
    				
\bibitem{snyder} H.S. Snyder, Phys. Rev. {\bf71}, 38 (1947).

\bibitem{nima1} N. Arkani-Hamed, S. Dimopoulos, and G. Dvali, Phys. Lett. B {\bf429}, 263 (1998).

\bibitem{nima2} I. Antoniadis, N. Arkani-Hamed, and G. Dvali, Phys. Lett. B {\bf436}, 257 (1998).

\bibitem{cms} S. Chatrchyan et al., Phys. Rev. D {\bf90}, 032008 (2014).

\bibitem{exp1} J. Cao, K. Hikasa, L. Wang, L. Wu, and J. M. Yang, Phys. Rev. D {\bf85}, 014025 (2012).

\bibitem{cakir} I.T. Cakir, O. Cakir, A. Senol, and A.T. Tasci, Acta Phys. Pol. B {\bf45}, 1947 (2012) .  				
\bibitem{phe1} Y. Wen, H. Qu, D. Yang, Q. Yan, Q. Li, and Y. Mao, J. High. Energy. Phys. {\bf03}, 025 (2015).

\bibitem{martin} G. Altarelli and D. Meloni, J. High. Energy. Phys. {\bf08}, 021 (2013).
		
\bibitem{bhowmick} M.G. Jackson and K. Schalm, 	Phys. Rev. Lett. {\bf108}, 111301 (2012).

\bibitem{susy1} T. Kobayashi and Y. Omura, J. High. Energy. Phys. {\bf02}, 114 (2015).

\bibitem{susy2} S. Raby, Eur. Phys. J. C {\bf59}, 223 (2009).

\bibitem{bran1} G.K. Leontaris, N.D. Tracas, N.D. Vlachos, and O. Korakianitis, Phys. Rev. D {\bf76}, 115009 (2007).

\bibitem{bran2} Y. Bu, Phys. Rev. D {\bf86}, 106005 (2012).

\bibitem{moyal1} J.E. Moyal, Proc. Cambridge Phil. Soc. {\bf45}, 99 (1949).

\bibitem{moyal2} M.R. Douglas and N.A. Nekrasov, Rev. Mod. Phys. {\bf73}, 977 (2002).

\bibitem{chaichian1} M. Chaichian, P. Pre\v{s}najdar, M.M. Sheikh-Jabbari, and A. Tureanu, Phys. Lett. B {\bf526}, 132 (2002).

\bibitem{hayakawa} M. Hayakawa, Phys. Lett. B {\bf748}, 394 (2000).

\bibitem{chaichian2} M. Chaichian, P. Pre\v{s}najdar, M.M. Sheikh-Jabbari, and A. Tureanu, Eur. Phys. J. C {\bf29}, 413 (2003).

\bibitem{jurco1} B. Jur\v{c}o, L. M\"{o}ller, S. Schraml, P. Schupp, and J. Wess, Eur. Phys. J. C {\bf21}, 383 (2001).

\bibitem{jurco2} B. Jur\v{c}o, L. M\"{o}ller, S. Schraml, P. Schupp, and J. Wess, Eur. Phys. J. C {\bf17}, 521 (2000).

\bibitem{calmet} X. Calmet, B. Jur\v{c}o, P. Schupp, J. Wess, and M. Wohlgenannt, Eur. Phys. J. C {\bf23}, 363 (2002).

\bibitem{melic1} B. Meli\'c, K. Passek-Kumeri\v{c}ki, J. Trampeti\'c, P. Schupp, and M. Wohlgenannt,  Eur. Phys. J. C {\bf42}, 483 (2005).

\bibitem{melic2} B. Meli\'c, K. Passek-Kumeri\v{c}ki, J. Trampeti\'c, P. Schupp, and M. Wohlgenannt,  Eur. Phys. J. C {\bf42}, 499 (2005).

\bibitem{garcia1} H. Garc\'ia-Compe\'an, O. Obreg\'{o}n, C. Ram\'irez and M. Sabido,  Phys. Rev. D {\bf68}, 044015 (2003).

\bibitem{aschieri} P. Aschieri, M. Dimitrijevic, P. Kulish, F. Lizzi and J. Wess, \textit{Noncommutaive Spacetime: Symmetries in noncommutative geometry and field theory} (Springer, Heidelberg, 2009), Lect. Note Phys. {\bf774}.

\bibitem{garcia2} H. Garc\'ia-Compe\'an, O. Obreg\'{o}n and R. Santos-Silva, Adv. Math. Phys.  {\bf15}, 845328 (2015).

\bibitem{behr} W. Behr, N.G. Deshpandeh, G. Duplan\v{c}i\'{c}, P. Schupp, J. Trampeti\'{c}, and J. Wess, Eur. Phys. J. C {\bf29}, 441 (2003).

\bibitem{buric2} M. Buric, D. Latas, V. Radovanovic, and J. Trampeti\'{c}, Phys. Rev. D {\bf75}, 097701 (2007).

\bibitem{xia} I. Mocioiu, M. Pospelov, and R.F. Lebed, Phys. Lett. B {\bf489}, 390 (2006).

\bibitem{betabi} S. Batebi, M. Haghighat, S. Tizchang, and H. Akafzadeh, Int. J. Mod. Phys. A  {\bf30}, 1550108 (2015).

\bibitem{ana2} A. Alboteanu, Ph.D Thesis, W\"{u}rzburg University, 2007.

\bibitem{ana3} A. Alboteanu, W. Kilian, and J. Reuter, J. High. Energy. Phys. {\bf11}, 010 (2008).

\bibitem{das1} P.K. Das, N.G. Deshpandeh, and G. Rajasekaran, Phys. Rev. D {\bf77}, 035010 (2008).

\bibitem{das2} P.K. Das, A. Prakash, and A. Mitra, Phys. Rev. D {\bf83}, 056002 (2011).

\bibitem{das3} A. Prakash, A. Mitra, and P.K. Das, Phys. Rev. D {\bf82}, 055020 (2010).

\bibitem{godf} S. Godfrey and M.A. Doncheski, Phys. Rev. D {\bf65}, 015005 (2001).

\bibitem{violation1} N. Seiberg, L. Sussskind, and N. Toumbas, J. High. Energy. Phys. {\bf06}, 044 (2000).

\bibitem{violation2} M. Chaichian, K. Nishijima, and A. Tureanu, Phys. Lett. B {\bf568}, 146 (2003).

\bibitem{pdg} K.A. Olive et al. (Particle Data Group), Chin. Phys. C {\bf 38}, 090001 (2014).

\end{thebibliography}
\end{document}